\documentclass{article}

\usepackage{arxiv}
\usepackage{listings}
\usepackage[utf8]{inputenc} % allow utf-8 input
\usepackage[T1]{fontenc}    % use 8-bit T1 fonts
\usepackage{url}            % simple URL typesetting
\usepackage{booktabs}       % professional-quality tables
\usepackage{amsfonts}       % blackboard math symbols
\usepackage{nicefrac}       % compact symbols for 1/2, etc.
\usepackage{microtype}      % microtypography
\usepackage{graphicx}
\usepackage{doi}
\usepackage{comment,amsmath,appendix}
\usepackage{soul}
\usepackage{amssymb}

\title{Multem 3: An updated and revised version of the program for transmission and band calculations of photonic crystals}

\author{ \href{https://orcid.org/0000-0002-4566-8093}{\includegraphics[scale=0.06]{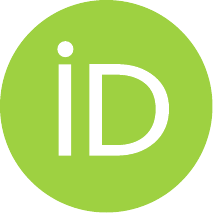}\hspace{1mm}Artem Shalev}\\
	School of Physics and Engineering, \\
        ITMO University, 191002, \\
        St. Petersburg, Russia \\
	\texttt{artem.shalev@metalab.ifmo.ru} \\
	\AND
	\href{https://orcid.org/0000-0001-6026-0777}{\includegraphics[scale=0.06]{orcid.pdf}\hspace{1mm}Konstantin Ladutenko} \\
	School of Physics and Engineering, \\
        ITMO University, 191002, \\
        St. Petersburg, Russia \\
        \AND
	\href{https://orcid.org/0000-0001-8789-3267}{\includegraphics[scale=0.06]{orcid.pdf}\hspace{1mm}Igor Lobanov} \\
	School of Physics and Engineering, \\
        ITMO University, 191002, \\
        St. Petersburg, Russia \\
        \AND
	\href{https://orcid.org/0000-0002-7832-7527}{\includegraphics[scale=0.06]{orcid.pdf}\hspace{1mm}Vassilios Yannopapas} \\
	National Technical University of Athens,\\
        NTUA Department of Physics
        \AND
	\href{https://orcid.org/0000-0003-4866-6517}{\includegraphics[scale=0.06]{orcid.pdf}\hspace{1mm}Alexander Moroz} \\
	\textit{Wave-scattering.com}
}

\date{}

\renewcommand{\shorttitle}{}

\hypersetup{
pdftitle={A template for the arxiv style},
pdfsubject={q-bio.NC, q-bio.QM},
pdfauthor={David S.~Hippocampus, Elias D.~Striatum},
pdfkeywords={First keyword, Second keyword, More},
}

\begin{document}
\maketitle

\begin{abstract}
We present here Multem 3, an updated and revised version of Multem 2,
which syntax has been upgraded to Fortran 2018, with the source code being divided into modules. Multem 3 is equipped with LAPACK, the state-of-the art
Faddeeva complex error function routine, and the Bessel function package AMOS. The amendments significantly improve both the speed, convergence, and precision of Multem 2. Increased stability allows to freely increase the cut-off value LMAX on the number of spherical vector wave functions and the cut-off value RMAX controlling the maximal length of reciprocal vectors taken into consideration. An immediate bonus is that 
Multem 3 can be reliably used to describe bound states in the continuum (BICs).  To ensure convergence of the layer coupling scheme, it appears that appreciably larger values of convergence paramaters  LMAX and RMAX are required than those reported in numerous published work in the past using Multem 2.
We hope that Multem 3 will become a reliable and fast alternative to generic commercial software, such as COMSOL Multiphysics, CST Microwave Studio, or Ansys HFSS, and that it will become the code of choice for various optimization tasks for a large number of research groups. The improvements concern the core part of Multem 2, which is common to the extensions of Multem 2 for acoustic and elastic multiple scattering and to the original layer-Kohn-Korringa-Rostocker (LKKR) code. Therefore, the enhancements presented here can be readily applied to the above codes as well.
\end{abstract}

% keywords can be removed

\noindent {\bf NEW VERSION PROGRAM SUMMARY}
\begin{small}
\\
\noindent
{\em Program Title:} Multem 3                                          \\
{\em CPC Library link to program files:} (to be added by Technical Editor) \\
{\em Developer's repository link:} \href{https://github.com/ArtyomShalev/Multem-3}{Multem-3} \\
{\em Code Ocean capsule:} (to be added by Technical Editor)\\
{\em Licensing provisions:} MIT  \\
{\em Programming language:} Fortran 2018                                   \\
{\em Journal reference of previous version:} N. Stefanou, V. Yannopapas, A. Modinos Computer Physics Communications 132 (2000) 189           \\
  %Only required for a New Version summary, otherwise leave out.
{\em Does the new version supersede the previous version?:} yes  \\
  %Only required for a New Version summary, otherwise leave out.
{\em Reasons for the new version:} The linear algebra part and the standard of the programming language are outdated. More precise algorithms for calculating special functions (Faddeeva complex error function and Bessel functions) have been introduced.
\\
  %Only required for a New Version summary, otherwise leave out.

\noindent {\em Summary of revisions:}

\begin{itemize}

\item[(i)] \textit{Code refactoring}.
Language standard was updated to Fortran 2018. The source code was divided into modules with CMake building system as highlighted in the flowchart of Fig. \ref{fig:modules}.
%%%%%%%%%%%%%%%%%%%%%
\begin{figure}[!htb]
\includegraphics[width=\textwidth]{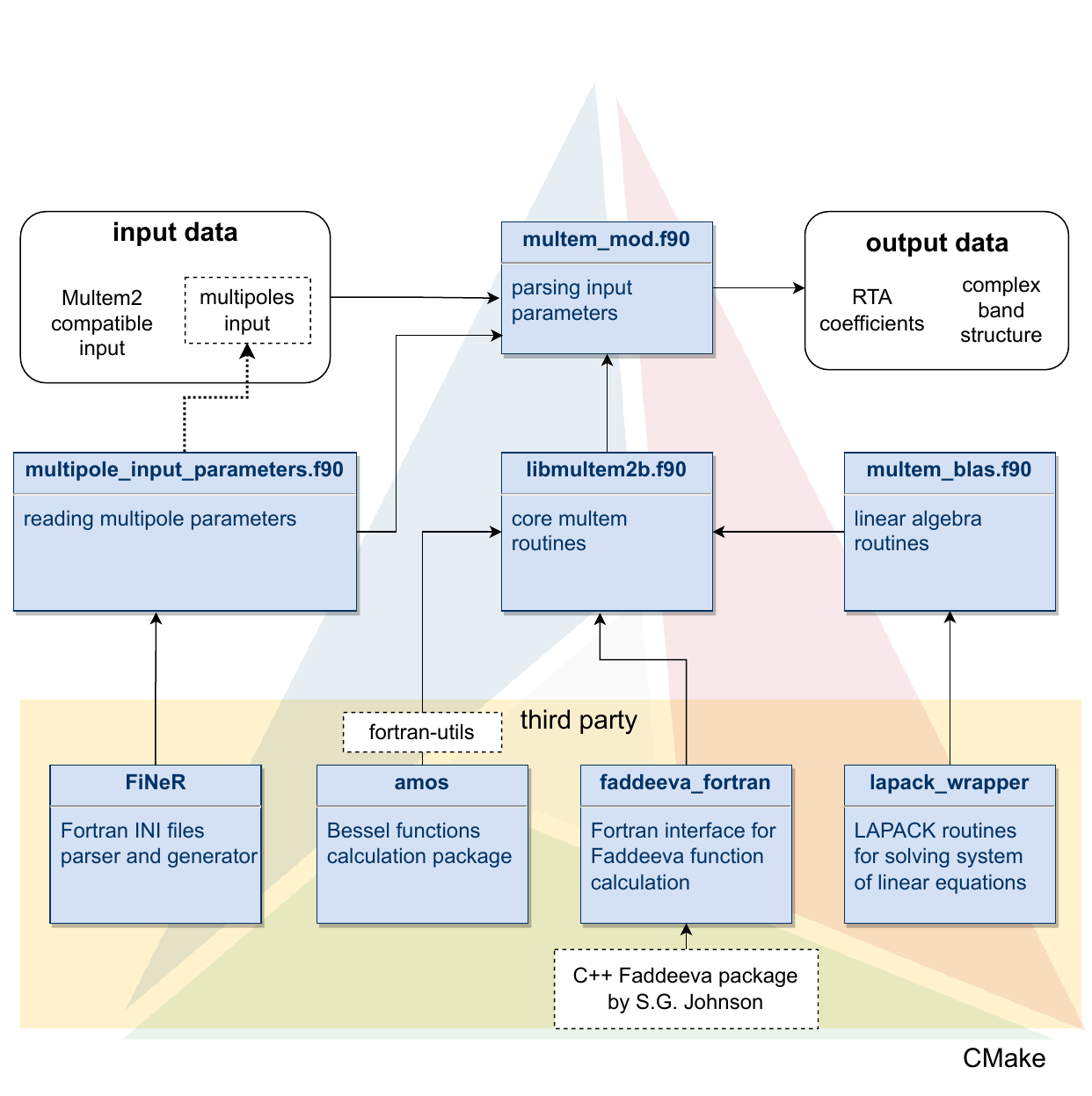}
\caption{Flowchart of modular structure of Multem 3}
\label{fig:modules}
\end{figure}
%%%%%%%%%%%%%%%%%%%%%

\item[(ii)]  \textit{Transition to LAPACK}.
Original routines ZGE and ZSU were changed to the LAPACK routines zgetrf and zgetrs,  resulting in a significant reduction of more than three orders of magnitude in the numerical errors that occur when solving systems of linear equations. 

\item[(iii)] \textit{Augmenting the implementation of the Faddeeva complex error function}.
Pendry's routine CERF~\cite{Pendry:LEED:1974_PSS} calculating the Faddeeva complex error function was substituted by the C++ SciPy code by Johnson~\cite{ScipyFaddeeva_PSS,FaddeevaMIT_PSS}. 
The latter shows significantly improved results, with the relative error limited to the range between $1.69\times 10^{-14}$ and $2.67\times 10^{-13}$ for the real and imaginary parts of the function values, respectively, for the arguments from $10^{-8}$ to $10^{8}$. 

\item[(iv)] \textit{Using the Bessel function package AMOS}.
Original routines for Bessel functions calculation, used to calculate the T-matrix of a single sphere in routine TMTRX, were substituted by the routines originally written by Donald E. Amos~\cite{Amos:614:1986_PSS}, which can be found at~\cite{AmosLib:2002_PSS}. Interfaces~\cite{FortranUtils_PSS} 
were used to link the code with the  original Amos routines. The benefit of this change was not directly evaluated. Nevertheless, the Amos library provides 
in our experience the most accurate results among all the libraries used to compute Bessel functions. 

\item[(v)] \textit{Increased stability allowing to freely increase the cut-off value LMAX on the number of spherical vector wave functions}.
LMAX is one of the most important convergence parameters of Multem. Nevertheless, any change of LMAX was very inconvenient as it required changing its value in almost any original routine. Furthermore, an upper cut-off value LMAXD was imposed on allowable LMAX values. Any attempt to increase LMAX above a preset value of LMAXD=7 required changing LMAX and LMAXD, together with additional dependent parameters NELMD and NDEND (cf. Table \ref{tab:LMAX} below). Contrary to that, Multem 3 allows calculations to be performed with any value of LMAX set at a single code location.

\item[(vi)] \textit{Increased stability allowing to freely increase the cut-off value RMAX controlling the maximal length of reciprocal vectors taken into consideration}.
RMAX is the second  of the main convergence parameters of Multem. Similarly to LMAX, varying RMAX required adjusting parameters in almost all original routines. Furthermore, an upper cut-off value RMAXD was imposed on allowable RMAX values. Any attempt to increase RMAX above a preset value of RMAXD required changing both RMAX and RMAXD, together with additional dependent parameter IGD (cf. Table \ref{tab:RMAX} below). 
Contrary to Multem 2, Multem 3 allows calculations to be performed with much larger values of RMAX which is now set at a single code location.

\item[(vii)] \textit{Selective multipole expansion}. 
Multem 3 has been extended by a selective multipole expansion option in order to be able to isolate the contribution of any given multipole. This allows one to perform a selective multipole analysis of a resonant 2D structure. We have adapted the routines SETUP and PCSLAB, which are used to construct the coefficient matrices for linear systems of equations and to calculate the transmission/reflection matrices for a plane of spheres, accordingly.

\end{itemize}

\noindent {\em Nature of problem:}\\
  Calculation of the transmission, reflection and absorption coefficients, and complex band structure of a photonic crystal slab. \\

\noindent {\em Solution method:}\\
  Solution of Maxwell's equations in combination with {\em ab-initio} multiple-scattering theory for 2D periodic systems and a layer-by-layer coupling scheme. \\
  
\noindent {\em Additional comments including restrictions and unusual features:}\\
  As its predecessor, Multem 3 can work only with layers of non-overlapping spheres arranged into simple Bravais lattice.
  By default, the underlying Bravais lattice has to be same in each plane of spheres (i.e. it is not possible to stack a triangular lattice on top of a square lattice, but it is possible to arrange different spheres (e.g. of different radius or of material) on the same underlying lattice in different planes). Additionally, an interface between planes may not cut any of the spheres.

\end{small}

%% main text

\vspace*{0.4in}

\section{Introduction}
\label{sec:intro}
%%%%%%%%%%%%%%%%%%%%%%%%%
Spherical scatterers arranged in periodic arrays have been studied numerically for at least a century~\cite{Rayleigh:1892, Darwin:1914:PartI, Darwin:1914:PartII}.
To determine reflection (R), transmission (T), and absorption (A) of a periodic structure, it proved very efficient to consider the structure as a stack of layers formed by infinite two-dimensional (2D) arrays of scatterers and to apply the general {\em ab-initio} multiple scattering theory only to the individual 2D layers~\cite{Rayleigh:1892, Darwin:1914:PartI, Darwin:1914:PartII}. In doing so, multipole expansion is used to solve the corresponding 2D  {\em ab-initio} multiple scattering problem. 
Physical quantities are expanded in conventional scalar or vector spherical harmonics denoted by orbital, $\ell$, and magnetic, $m$, angular numbers, where $\ell\in [1, \infty]$ and $m \in [-\ell, \ell]$. To perform calculations,  a cut-off value LMAX is introduced for $\ell$ and convergence with increasing cut-off value of LMAX is investigated. An intermediate outcome is a scattering matrix of each individual layer in the basis of reciprocal lattice vectors. 
The final solution is then obtained by coupling the individual scattering matrices using an ingenious recursive layer-by-layer coupling scheme~\cite{Pendry:LEED:1974,Pendry:1973}, which, in the case of identical layers, enables one to double the number of layers at each step. The number of diffraction orders ${\bf k}_n$, or reciprocal vectors, considered in the calculation controls the coupling of the respective planes and layers of spheres. In addition to ensuring convergence in LMAX, one must ensure that the coupling is correct. The latter is done by choosing an appropriate value RMAX, which is a cutoff on the length of reciprocal vectors ${\bf k}_n$ taken into account for diffraction orders.
Splitting the initial intrinsic 3D multiple-scattering problem into a 2D multiple-scattering problem in combination with the layer coupling scheme proved to be numerically very efficient.
The above algorithm has been employed in a number of state-of-the-art multiple scattering computational methods, such as low-energy electron diffraction (LEED) ~\cite{Pendry:LEED:1974} and its underlying Kohn-Korringa-Rostocker (LKKR) method~\cite{McL:KKR:1990, McL:KKR1:1990}, as well as their extensions for electromagnetic~\cite{multem1, multem2}, acoustic~\cite{Kafesaki:1999}, and elastic scattering~\cite{Kafesaki:2000,Liu:2000}. A number of breakthroughs were required to arrive at the state-of-the-art computational methods for multiple scattering.

\begin{itemize}
 \item 
The first breakthrough required an extension of the Ewald summation~\cite{Ewald1921} to the case when summation is performed in smaller dimension than that of embedding space (e.g. over 2D array embedded in three-dimensions~\cite{Kambe:1967:I, Kambe:1967:II, Kambe:1968, Moroz:2006}).

\item The second one required an efficient and stable algorithm enabling one to couple different 2D arrays of scatterers along the stacking directions and thereby generate more complex structures~\cite{Pendry:LEED:1974,Pendry:1973}.
\end{itemize}

These two pillars are the common foundation of all successful multi-scattering computational schemes~\cite{Pendry:LEED:1974,McL:KKR:1990, McL:KKR1:1990,multem1, multem2,Kafesaki:1999,Kafesaki:2000,Liu:2000}. Recently published software {\it treams} for electromagnetic scattering calculations~\cite{Beutel:treams:2023} based on the T-matrix method seems to also employ Kambe's lattice summation~\cite{Beutel:2D3Dsums:2023, Beutel:biperiodic:2021} and comprising its extensions as reviewed in~ \cite{Moroz:2006,Moroz:2001}. 

An exhaustive definition and description of all relevant parameters, together with detailed algorithm description can be found (i) in classical textbook by Pendry~\cite{Pendry:LEED:1974} and (ii) in a long write-up of Multem 1~\cite{multem1} that we strongly recommend for further reading.
In computational terms, all of the above codes have inherited as their core the original Pendry Fortran 77 code for scalar waves written almost 50 years ago in 1974, the listing of which is presented in the appendix of Pendry's monograph~\cite{Pendry:LEED:1974}. The core part includes, among other, (1) the lattice sum part which implements Kambe's summation ~\cite{Kambe:1967:II} in the routine XMAT~\cite[Appendix C]{Pendry:LEED:1974}, (2) all linear algebra routines, (3) all required special (e.g. Bessel, Faddeeva, and Legendre) functions. 

The focus of present work is on multem codes~\cite{multem1,multem2} which solve the Maxwell's equations using {\em ab-initio} multiple-scattering theory~\cite{Pendry:LEED:1974,Butler:1992}. It turns out that the core part of Multem 2~\cite{multem2}, common to also other related codes~\cite{Pendry:LEED:1974,McL:KKR:1990, McL:KKR1:1990,Kafesaki:1999,Kafesaki:2000,Liu:2000}, was never updated since the time of its writing in 1974~\cite{Pendry:LEED:1974}.
The linear algebra part has remained largely outdated and has been for decades begging for a major overhaul. Note that at the time the program core part was written~\cite{Pendry:LEED:1974}, the so-called double precision was hardly available. 
Not surprisingly, many users have experienced various degree of fragility when running the code. While an apparent convergence of results was observed with gradual increase of the angular momentum cutoff value LMAX, say from $3$ to $5$, further increases in LMAX to $7$ or higher values may have resulted in an abrupt loss of convergence. As larger and larger matrices were included in the linear algebra operation, the condition number of matrices increased and eventually convergence was destroyed. At other occasions, convergence to nonphysical results was encountered.
Not surprisingly, the current Multem 2~\cite{multem2} is, as a rule, not suitable to provide a reliable description of the so-called bound states in the continuum (BICs). Since initial experimental observation of BICs in photonic crystals slabs in 2013~\cite{HsuBIC2013}, the BICs have become an important direction in photonic research due to BICs strong spatial localization and high quality factors. 
The demands on numerical precision for a reliable description of structures supporting BICs are simply too high for the currently available versions of Multem 1~\cite{multem1} and Multem 2~\cite{multem2}. It was mainly the above mentioned failure 
of Multem 2~\cite{multem2} that motivated us to look for its improvements. 

For the sake of notation, we refer to the new version as Multem 3, given that the presently available version of the code is called~Multem 2~\cite{multem1, multem2}.
Multem 3 brings many substantial improvements. First, its syntax has been upgraded from Fortran 77 to Fortran 2018, with the source code being divided into modules (see the flowchart of Fig. \ref{fig:modules}). On numerical side, the linear algebra part of Multem 2 has been made up-to-date by replacing original routines with the LAPACK routines for solving relevant systems of linear equations. Note that this upgrade was still not enough for providing a reliable description of the BICs. Surprisingly enough, the crucial point was upgrading Kambe's summation~\cite{Kambe:1967:II} in routine XMAT
involving the lattice sum part $D_{LM}^{(1)}$~\cite{Kambe:1967:II,Moroz:2006}). The upgrade consists in that the routine CERF \cite[pp. 342-4]{Pendry:LEED:1974}
generating an incomplete gamma function (whose special cases are the error function and the complementary error function~\cite{AbramowitzStegun, NumRecipes2007}) has been replaced by the state-of-the art routine for the so-called {\em Faddeeva function} (also known as the complex complementary error function of complex argument).

Additionally, Multem 3 features (i) code refactoring, and (ii) adding a selective multipole expansion option. 
The extensions allow for both speed and accuracy improvements in Multem 3, thereby enabling  reliable description of BICs. We hope that Multem 3 will become a reliable and fast alternative to commercial software such as COMSOL Multiphysics, CST Microwave Studio, or Ansys HFSS. 
Commercial codes have the advantage of being applicable to a wider range of problems, but their generality also brings disadvantages. Not surprisingly, general purpose codes can be much slower than highly specialized codes optimized for a specific task. This was already the case when commercial codes were compared with earlier versions of \textit{Multem 1}~\cite{multem1} and Multem 2~\cite{multem2}. The gain in precision and computational speed will be even more significant in the current version Multem 3. We hope that the updated open-source code Multem 3 has the potential to become the code of choice for a large number of research groups for various optimization tasks.

In Sec.~\ref{sec:method}, we describe the execution path for calculating transmittance spectrum of 2D arrays of dielectric spheres using the names of input variables and subroutines from~\cite{multem1}. In Sec.~\ref{sec:comparison}, we show how calculation accuracy has increased with several modifications of Multem 2. 

In Sec.~\ref{sec:lapack} we demonstrate on an example the effect of including LAPACK routines on the code precision.

In Sec.~\ref{section:lmax} we show improvements of Multem 3 over Multem 2 in the code stability and precision when varying the in-plane convergence parameter LMAX, which, contrary to Multem 2, can be selected freely in Multem 3.

In Sec.~\ref{sec:RMAX} we show improvements of Multem 3 over Multem 2 in the code stability and precision when varying the inter-plane convergence parameter RMAX.
On an example of reflectance spectra it will be illustrated that Multem 2 may converge with increasing RMAX to {\em nonphysical} results (calculated R exceeds physical bound of $R\le 1$), whereas converged reflectance spectra calculated by Multem 3 safely obey the physical bound of $R\le 1$ over all range considered and never exceed the horizontal line $R=1$. 

In Sec.~\ref{sec:faddeeva} we show the effect of implementing the state-of-the art Faddeeva function package on the code performance.

In Sec.~\ref{sec:multipole_expansion}, we describe the selective multipole expansion implementation of Multem 3 which allows one to easily isolate the contribution of any particular multipole.  

In Sec.~\ref{sec:limitations} we discuss current limitations of Multem 3 and suggest its further potential improvements. We then conclude with Sec.~\ref{sec:concl}.

\section{Code overview}
\label{sec:method}
\begin{figure}[h]
\includegraphics[width=\textwidth]{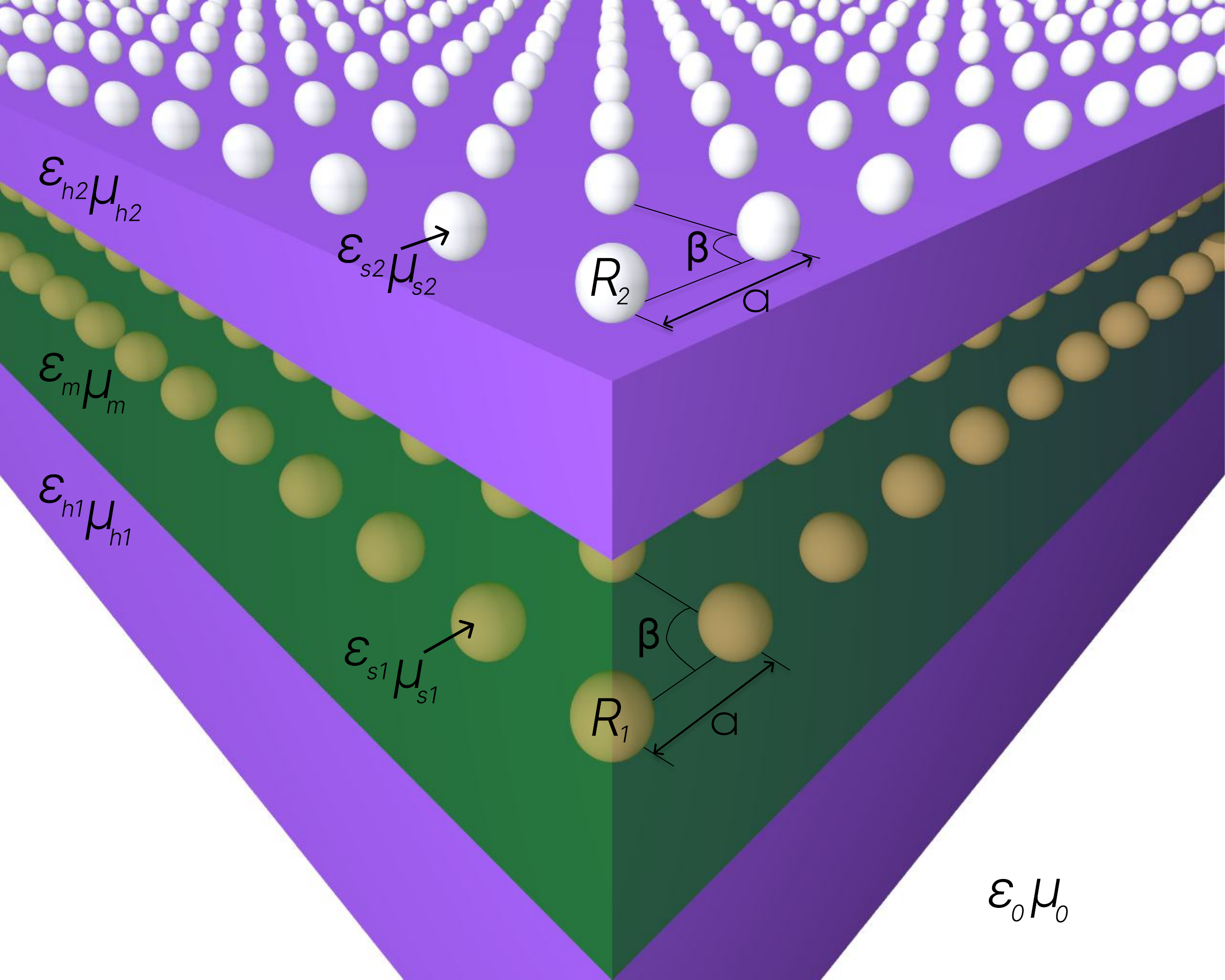}
\caption{
Typical infinite 2D periodic structure consisting of a finite number of planes. Each plane can be either homogeneous ($\varepsilon_h, \mu_h$) plate or an infinite 2D lattice of non-overlapping spheres ($\varepsilon_s, \mu_s$) embedded in a homogeneous medium ($\varepsilon_m, \mu_m$). Spheres are arranged in a lattice defined by the lattice period $a$ and the angle between the primitive vectors of the 2D lattice $\beta$.}
\label{fig:typical_system} 
\end{figure}

%%%%%%%%%%%%%%%%%%%%%%%%%
Multem codes have been designed to deal with very complex superstructures going well beyond common crystallographic lattices~\cite{multem1,multem2}. Elementary scattering and diffractive objects of the code are called {\em planes}. The code allows for large versatility in that a plane can be either (i) a homogeneous (dielectric or metal) plate or (ii) a plane of non-overlapping spheres arranged in a simple Bravais lattice embedded in a homogeneous medium (Fig.~\ref{fig:typical_system}). This allows one to study an array of spheres, or a more complex photonic crystal slab, on a substrate. Alternatively, one can consider an array of spheres, or a more complex photonic crystal slab, sandwiched between two different substrates, with further planes allowed to be arranged on the opposite sides of the substrates. When stacking planes on top of each other, otherwise identical neighbouring planes are considered to be different if the planes are shifted relatively to each other in the transverse direction. It is expedient to identify repeating patterns when stacking the planes (e.g. ABCABCABC\ldots stacking) and to identify the repeating patterns as {\em layers} (e.g. layer ABC). For the sake of illustration, a face-centered cubic (fcc) lattice stacking in the (111) crystal direction involves just such a ABCABCABC\ldots stacking of identical infinite 2D triangular lattices of spheres. An example of trivial AAAA\ldots stacking is furnished by stacking 2D square lattices of simple cubic (sc) lattice in the (100) crystal direction.
The code allows for subsequent stacking of different layers and planes. Repeating patterns in such a stacking are identified as {\em slices}.  Identifying repeating pattern is expedient in that the code can, for example, calculate a layer scattering matrix, by applying the layer-by-layer coupling scheme determine the scattering matrix of a two-layer structure, and by subsequent iteration of the layer-by-layer coupling scheme couple two two-layer structures thereby arriving at a four-layer structure, etc., thus {\em doubling} the number of layers at each subsequent iteration step ~\cite{Pendry:1973}.
One can configure the number of slices, their thickness, complex permittivity $\varepsilon$ and permeability $\mu$ of spheres and embedding media, radii of spheres, and the offset between the layers consisting of spheres. Technical details of this was summarized neatly in original long write-up of the code~\cite{multem1}.
By default, the underlying Bravais lattice has to be same in each plane of spheres (i.e. it is not possible to stack a triangular lattice on top of a square lattice, but it is possible to arrange different spheres (e.g. of different radius or of material) on the same underlying lattice in different planes). Additionally, an interface between planes may not cut any of the spheres. A standard version of the code does not allow also simulate a system with two lattices of the same kind but with different parameters, for example, two square lattices with different pitch.
This is because the layer-by-layer coupling scheme involves matrices indexed by the reciprocal vectors. Changing any of lattice kind or basis lattice vector implies different reciprocal vectors. The latter prevents applying the layer-by-layer coupling algorithm, because relevant scattering square matrices would be expressed in different bases. Contrary to that, a displacement between identical lattices is a common option in the code, which is specified by the vector DL and DR~\cite{multem1,multem2}. Such a displacement is intrinsic for common naturally occurring crystals, such as, for example, an fcc lattice stacked along the (111) crystal direction.

The program can be executed in two modes. The first one allows to calculate experimentally measured quantities such as the reflection (R), transmission (T), and absorption (A) of an incident plane wave by a photonic structure. The second one allows one to calculate the complex 2D photonic band structure of an infinite periodic crystal. 
%%%%%%%%%%%%%%%%%%%%%%%%%
%%%%%%%%%%%%%%%%%%%%%%%%%
For the sake of comparison, and for readers sake, we preserve the notation of original multem versions Multem 1~\cite{multem1} and Multem 2~\cite{multem2}. General steps to calculate the RTA coefficients for a given system are: 
\begin{enumerate}

\item Construct direct and reciprocal lattice vectors using subroutine LAT2D. 

\item Expand an incident plane wave described by polarization vector EIN and wave vector magnitude KAPIN into spherical vector wave functions (SVWF) using subroutines PLW and SPHRM4.

\item The system of linear equations, which connects the incident field, scattered field, scattering properties of the individual sphere, and multiple-scattering between the spheres within single layer, is constructed using subroutine PCSLAB (see Eq.~\ref{eq:main} in Sec~\ref{sec:multipole_expansion} and Eq. 36 in~\cite{multem1}). $\mathbf{T}$-matrices of the individual sphere are calculated by subroutine TMTRX. The multiple-scattering of spheres in plane is taken into account by the so-called Ewald-Kambe lattice summation~\cite{Ewald1921, Kambe:1967:I} that is implemented in subroutine XMAT. Then $\mathbf{\Omega}$-matrices are calculated and used in the left-hand side of the system (cf. Eq.~\ref{eq:main} below) constructed by subroutine SETUP. The right-hand side of the system is constructed from expansion coefficients of the incident field (calculated at the previous step) and $\mathbf{T}$-matrices.
  
\item Find the solution of the mentioned system in terms of expansion coefficients of the scattered field which is also a part of the subroutine PCSLAB.

\item Express scattered field off an array of spheres as a sum of plane waves and construct transmission and reflection matrices $\mathbf{M}$. These matrices connect incident and scattered plane waves for the slice (subroutines PCSLAB and DLMKG).
  
\item Calculate scattering $\mathbf{Q}$-matrices~\cite{Pendry:1973} for the whole layer consisting of a number of slices by multiplying $\mathbf{M}$-matrices with the appropriate phase factors (subroutine PCSLAB). 
   
\item Calculate the transmission, reflection and absorption coefficients of a finite slab by defining transmitted and reflected electric field corresponding to the given incident field and integrating the Poynting vector over the $xy$-plane taking the time average over a incident field oscillation period (subroutine SCAT).
  
\end{enumerate}
%%%%%%%%%%%%%%%
For the sake of illustration of Multem 3 capabilities, we shall use throughout our subsequent presentation two examples taken from the current literature. 

The first one involves triangular 2D lattice consisting of high-index ($\varepsilon = 15$) dielectric lossless spheres embedded in vacuum, which set-up is shown in Fig. 1 of Ref.~\cite{Bulgakov:19}. In multem language, the system corresponds to the transmittance spectra (KSCAN=1) calculated for the TE polarized plane wave (POLAR=S, FEIN=0) incident on a triangular (FAB=60) lattice with lattice period $a$ (ALPHA=1, BETA=1) consisting of dielectric spheres (EPSSPH=15, MUSPH=1) of radius $R=0.4705a$ (S=0.4705), embedded in vacuum (KEMB=0, EPSMED=1, MUMED=1). The system contains only one plane (NCOMP=1, NUNIT=1, NPLAN=1, NLAYER=1) of spherical particles (IT=2). The angle of incidence is set by the in-plane components of an incident wave vector $(k_x,k_y)$ stored in array AK (KTYPE=2). The example of Ref.~\cite{Bulgakov:19} was chosen, because it exhibits an optical BICs in the TE wave spectrum at normal incidence. The BIC corresponds to an exceptional point located at the Brillouin zone point $\Gamma$ corresponding to the Bloch vector $(k_x,k_y)=0$.

The second example comprises involves a single 2D triangular lattice of Au spheres of Ref.~\cite{Swiecicki:17}.

\section{Comparison with Multem 2}
\label{sec:comparison}
%%%%%%%%%%%%%%%%%%%%%%%%%%%%%%%%%%%%%%%%%%%%%%%%%%%%%%%
After the RTA coefficients are calculated, it is expedient to check the conservation of energy, which is expressed by the the constraint $R+T+A=1$. The latter will be used as a useful criterion for assessing the accuracy of calculations.  
It has been familiar to Multem 2 users that even when non-absorbing spheres with real material parameters were considered, in which case $A\equiv 0$, the code usually yielded $A=1 - (R+T)\ne 0$. The difference could be sometimes significant. 
In order to demonstrate the benefits of the improvements
in Multem 3 over Multem 2 mentioned in Sec.~\ref{sec:intro}, we recalculate the transmission spectra for the system considered by Bulgakov and Maksimov~\cite{Bulgakov:19} in order to demonstrate the effect of LAPACK improvement in section \ref{sec:lapack} and to investigate convergence with increasing values of LMAX in section \ref{section:lmax}. 
In sections~\ref{sec:RMAX}  we then illustrate our results by recalculating examples of Fig. 4, 7, 8 of Ref.~\cite{Swiecicki:17} involving a single 2D triangular lattice of Au spheres.
In Sec.~\ref{sec:faddeeva} we return back to the example of Ref.~\cite{Bulgakov:19} in order to demonstrate Multem 3 capability of dealing with BIC's.

\subsection{LAPACK}
\label{sec:lapack}
%%%%%%%%%%%%%%%%%%%%
The problem of finding scattering field coefficients is
in Multems reduced to solving systems of linear equations (see, e.g., Eq. (36) in Ref.~\cite{multem1}). Custom implementations of such routines can be source of errors, since the linear algebra calculations can be nontrivial. We have substituted the original ZGE, which computes an LU factorization, by LAPACK's routine ZGETRF, and the original ZSU, which solves a system of linear equations using the LU factorization, by the LAPACK's routine ZGETRS~\cite{LAPACKroutines}. 
%%%%%%%%%%%%%%%%%%%%%%
\begin{figure}[!htbp]
\includegraphics[width=\textwidth]{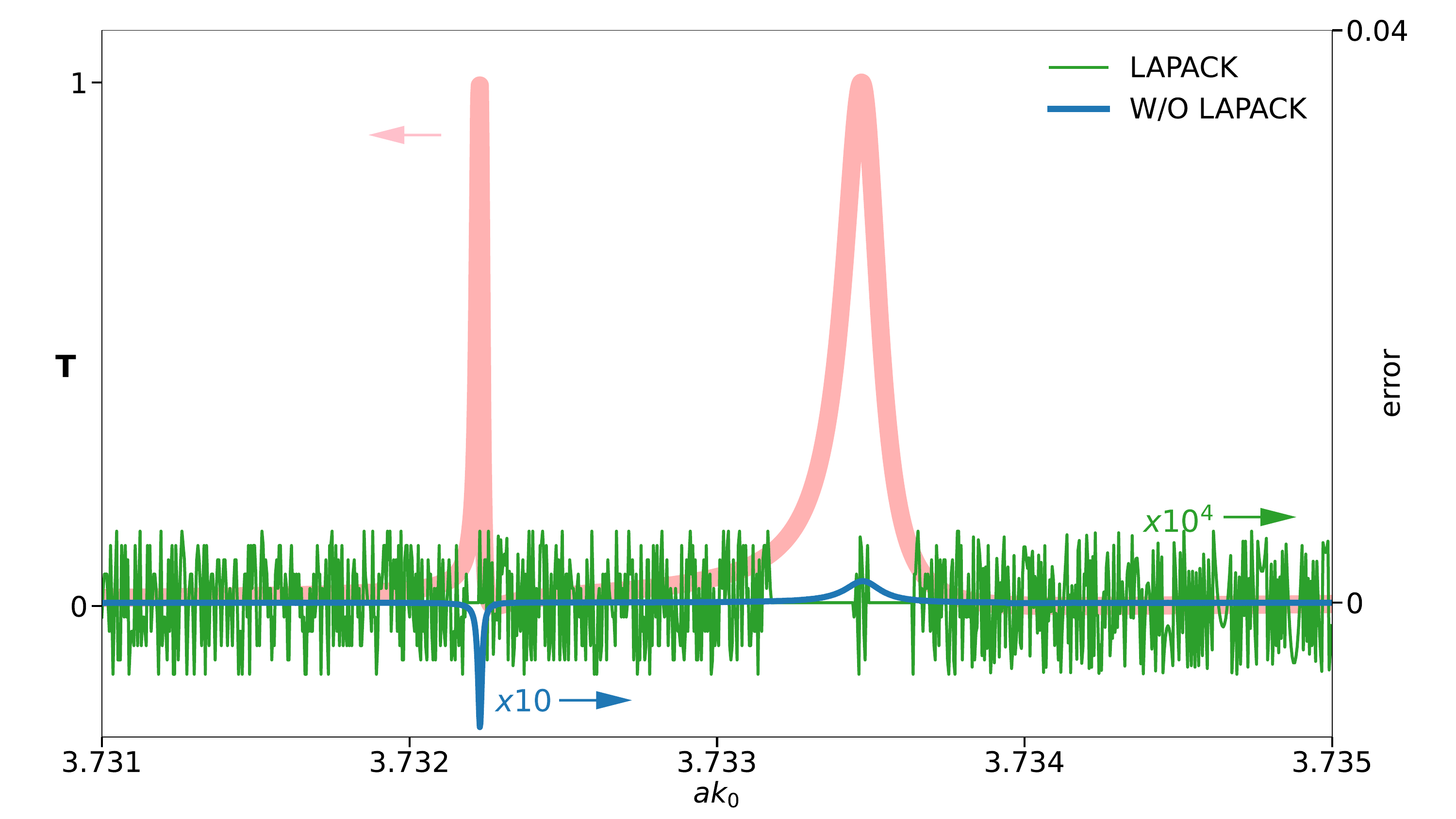}
\caption{The net effect of employing LAPACK in Multem 2 in the absence of any further improvement reported here with the cut-off values LMAX=7 and RMAX=16. Numerical error measured as a deviation of $A=1-R-T$ from zero obtained by Multem 2 with and without LAPACK is shown on the right axis. The error values obtained when using LAPACK had to be multiplied by $10^4$ in order to make them visible to naked eye. It takes 137 seconds of AMD®Ryzen 5 3500u time to calculate data for this plot (2000 data points for each spectra).}
\label{fig:main1}
\end{figure}
%%%%%%%%%%%%%%%%%%%%%%

To demonstrate the advantages of using LAPACK routines, Fig. \ref{fig:main1} shows in orange transmission spectrum for the setup presented by Bulgakov and Maksimov~\cite{Bulgakov:19} for the incident TE wave with $ak_x=0.01$ and $ak_y=0$ by means of Multem 2 with the LAPACK routines with the cut-off values LMAX=10 and RMAX=16. 
The right axis of Fig.~\ref{fig:main1} shows the error $= 1 - (R+T)$, calculated with Multem 2 with and without the LAPACK routines. Multem 2 with LAPACK routines shows $\lesssim |10^{-5}|$ numerical error with average numerical fluctuations over all frequency ranges and $\lesssim 6.98\times 10^{-8}$ numerical error over the resonance region. 
This clearly demonstrates that augmenting Multem 2 with LAPACK is very beneficial, as it reduces the numerical error of the constraint $R+T+A=1$ by more than three orders of magnitude compared to the original Multem 2.

%%%%%%%%%%%%%%%%%
\begin{figure}[!htbp]
\includegraphics[width=\textwidth,height=0.7\textheight]{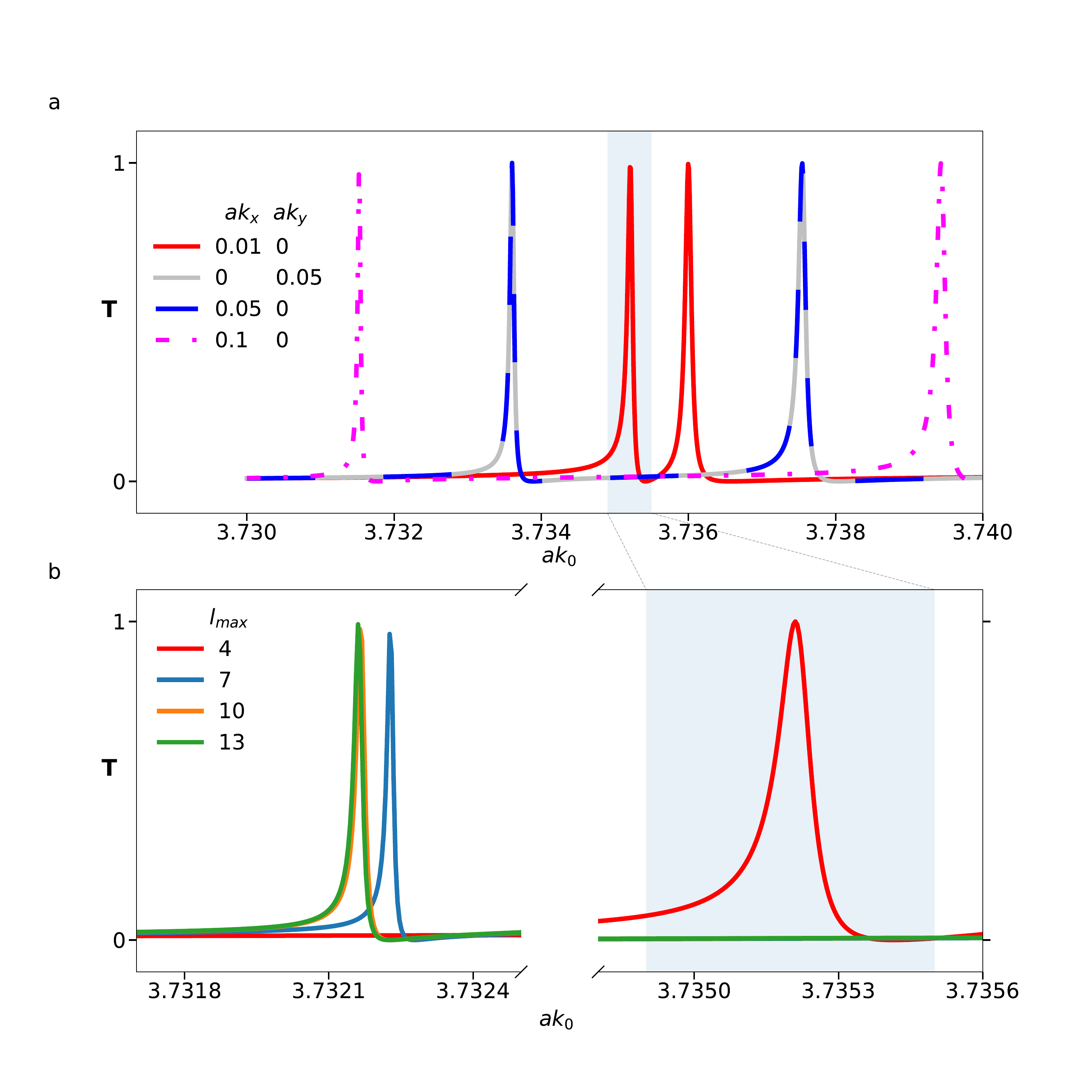}
\caption{(a) An illustrative example of reproducing the transmittance spectra of Fig. 3 of Ref.~\cite{Bulgakov:19} for high-index dielectric lossless spheres ($\varepsilon_s = 15$) arranged on a triangular 2D lattice and embedded in vacuum for different angles of incident wave characterized by $ak_x$ and $ak_y$ with LMAX=4.
Away from an exceptional point located at the Brillouin zone point $\Gamma$  (corresponding to $(k_x,k_y)=0$), the TE wave spectrum reveals two isolated Fano resonances whose positions are given by the real
parts of the Dirac cone eigenfrequencies and which distance increases with increasing $ak_x$ for fixed $ak_y=0$, and vice versa.  In the vicinity of the exceptional point, the Fano resonances merge into a single feature, which can only be resolved on a zoomed scale in $k_0$.
(b) Evolution of the transmittance spectra of the incident TE plane wave for $ak_x=0.01$ and $ak_y=0$ with increasing the cut-off value of LMAX. The grayed rectangle areas in (a) and (b) highlight the same peak obtained with LMAX$=4$. The results of Fig. 3 of Ref.~\cite{Bulgakov:19}, which were obtained with LMAX=4, are shown by solid red line. It takes 124 seconds of AMD®Ryzen 5 3500u time to calculate data for this plot (1000 data points for each spectra). All the plots were generated with convergence parameter RMAX=16.}
\label{fig:main2}
\end{figure}
%%%%%%%%%%%%%%%%%%%%%%

\subsection{The cut-off on the number of spherical vector wave functions LMAX}
\label{section:lmax}
%%%%%%%%%%%%%%%%%%%%%%%%%%%%%%%%%%%%%%%%%%%%%%%%%%%%%%%%%%%%%%%
The cut-off value LMAX imposed on the multipole expansion into spherical vector wave functions (SVWF) is one of the main convergence parameters (the other being RMAX) that influences the accuracy and calculation speed in Multem. By default, Multem 2 enables perform calculations with LMAX $\in [1, 7]$. For a number of application, one often needs larger values of LMAX~\cite{Moroz2002} (cf. also Fig.~\ref{fig:main2}).
Raising LMAX above 7 requires to increase correspondingly 
NELMD, the number of the Clebsh-Gordan coefficients, and NDEND (in XMAT). For the readers convenience, the Table below provides instruction in how the values of NELMD and NDEND have to be adjusted for a given LMAX.
%%%%%%%%%%%%%%%%%%%%%%
\begin{table}[h!]
 \centering  
 \caption{Varying the cut-off LMAX in Multem 2}
 \vspace*{0.3cm}
 \begin{tabular}{ccc}\toprule
  LMAX  &  NELMD  &   NDEND  \\
  \midrule 
 4 &  809   &    55 \\
 5   &1925    &   91 \\
 6   &4032   &   140 \\
 7    &    7680    &  204 \\
 8 & 13593    &  285 \\
 9   &    22693    &  385 \\
10   &    36124 & 506 \\
11    &   55226 & 650 \\
12   &    81809   &   819 \\
13  &    117677    & 1015 \\
14    &  165152    & 1240 \\ \bottomrule
\label{tab:LMAX}
\end{tabular}
\end{table}
LMAX as large as $14$ was employed when performing calculations for zinc-blende and diamond lattices of core-shell spheres~\cite{Moroz2002}. Nevertheless, varying the cut-off value of LMAX was rather cumbersome as it required adjusting parameter values in a great deal of routines. In contrast to Multem 2, Multem 3 allows to set arbitrary value of LMAX at a single place. 

%In Fig. \ref{fig:main}(a) numerical convergence was investigated with increasing LMAX. 
%Results of Ref.~\cite{Bulgakov:19} reproduced with LMAX=4 are shown by solid red line in \ref{fig:main}(b).

Fig.~\ref{fig:main2}(b) investigates LMAX convergence on the example 
of transmittance spectrum for the setup studied in Fig. 3 of Ref.~\cite{Bulgakov:19}. 
Although it was claimed in Ref.~\cite{Bulgakov:19} that the convergence has been checked by increasing the number of multipoles with no significant effect, LMAX$=4$ chosen in Ref.~\cite{Bulgakov:19} is clearly not enough to ensure convergence of numerical solution: there is $24.28$ full width at half maximum (FWHM) resonance position shift in the respective spectra obtained with LMAX$=4$ and LMAX$=5$. 
In their simulations, Bulgakov and Maksimov~\cite{Bulgakov:19} used closely related LKKR method of Ohtaka~\cite{Ohtaka:1979,Ohtaka:1980}, which has its roots also in the LEED. It is just an alternative implementation of the same underlying theory. Therefore, there is no reason to assume that the LKKR method of Ohtaka~\cite{Ohtaka:1979,Ohtaka:1980} is behaving differently.
Further increasing LMAX$=7$ is still not enough ($3.31$ FWHM offset). As demonstrated in Fig.~\ref{fig:main2}(b), at least LMAX$=10$ (resulting in $0.04$ FWHM offset) should be taken in order to get a satisfactorily accuracy for most of applications. Note in passing that the value of LMAX$=10$ is well beyond default Multem 2 options. Contrary to that, essentially any value of LMAX can be selected in Multem 3.

%Obviously, the results of Ref.~\cite{Bulgakov:19} were not fully converged.
%The convergence has been checked by increasing the number of multipoles with no significant effect.

\subsection{The cut-off on the number of diffraction orders RMAX}
\label{sec:RMAX}
%%%%%%%%%%%%%%%%%%%%%%%%%%%%%%%%%%%%%%%%%%%%%%%%%%%%%%%%%%%%%%%%%%
A plane wave $\sim e^{i{\bf k}\cdot{\bf r}}$ incident on a 
scattering plane of identical scatterers arranged regularly 
on a lattice $\Lambda$ is, in general, diffracted (transmitted) 
to a wave with a wave vector ${\bf K}_n^-$ (${\bf K}_n^+$), 
where ${\bf K}_{n}^\pm = \left({\bf k}_\parallel+{\bf k}_n,  
K_{\perp n}^\pm \right)$, 
\begin{equation}
K_{\perp n}^\pm = \pm K_{\perp n}=\left\{
\begin{array}{cc}
\pm [\sigma^2-|{\bf k}_\parallel+{\bf k}_n|^2]^{1/2},
      &\sigma^2 \geq |{\bf k}_\parallel+{\bf k}_n|^2\\
\pm i[|{\bf k}_\parallel+{\bf k}_n|^2-\sigma^2]^{1/2},
      &\sigma^2<|{\bf k}_\parallel+{\bf k}_n|^2,
\end{array}\right.
\label{Kkn}
\end{equation}
${\bf k}_\parallel$ is the projection of incident wave vector ${\bf k}$
on the scattering plane, ${\bf k}_n\in \Lambda^*$ are different Bragg diffraction orders, 
i.e. elements of the dual, or reciprocal, lattice $\Lambda^*$~\cite{Pendry:LEED:1974,multem1,multem2,Kambe:1967:II,Moroz:2006}.
Parameter $\sigma$ stands for the magnitude of $|{\bf k}|=|{\bf K}_{n}^\pm|=\sigma$. 
In the above definition, the normal projection $K_{\perp n}$ 
can be either real or imaginary. In the case of real 
$K_{\perp n}$ we speak of a {\em propagating wave}, and 
in the case of imaginary $K_{\perp n}$ of an {\em evanescent wave}. 

The number of diffraction orders ${\bf k}_n$, or reciprocal vectors, taken into account in the computation
controls the coupling of respective planes and layers of spheres. In addition to ensuring convergence in LMAX, which controls convergence in the  planes and layers of scatterers, 
one has to ensure that the coupling is correct. The latter is done by selecting an appropriate value RMAX, which is a cutoff on the length of reciprocal vectors ${\bf k}_n$ taken into account for diffraction orders. 
As a rule of thumb, one includes all diffraction orders for which $K_{\perp n}$ is real and includes a few evanescent orders for which $K_{\perp n}$ is purely imaginary~\cite{Pendry:LEED:1974,multem1,multem2,Kambe:1967:II,Moroz:2006}. Similarly to varying the in-plane convergence parameter LMAX (cf. Table \ref{tab:LMAX}), there is a dependent parameter value IGD, which value has to be adjusted when varying the inter-plane convergence parameter RMAX. This depends on the lattice type as summarized in the Table below for a face-centered cubic (fcc) and a simple cubic (sc) lattices.
%%%%%%%%%%%%%%%%
\begin{figure}[h!]
  \includegraphics[width=\textwidth]{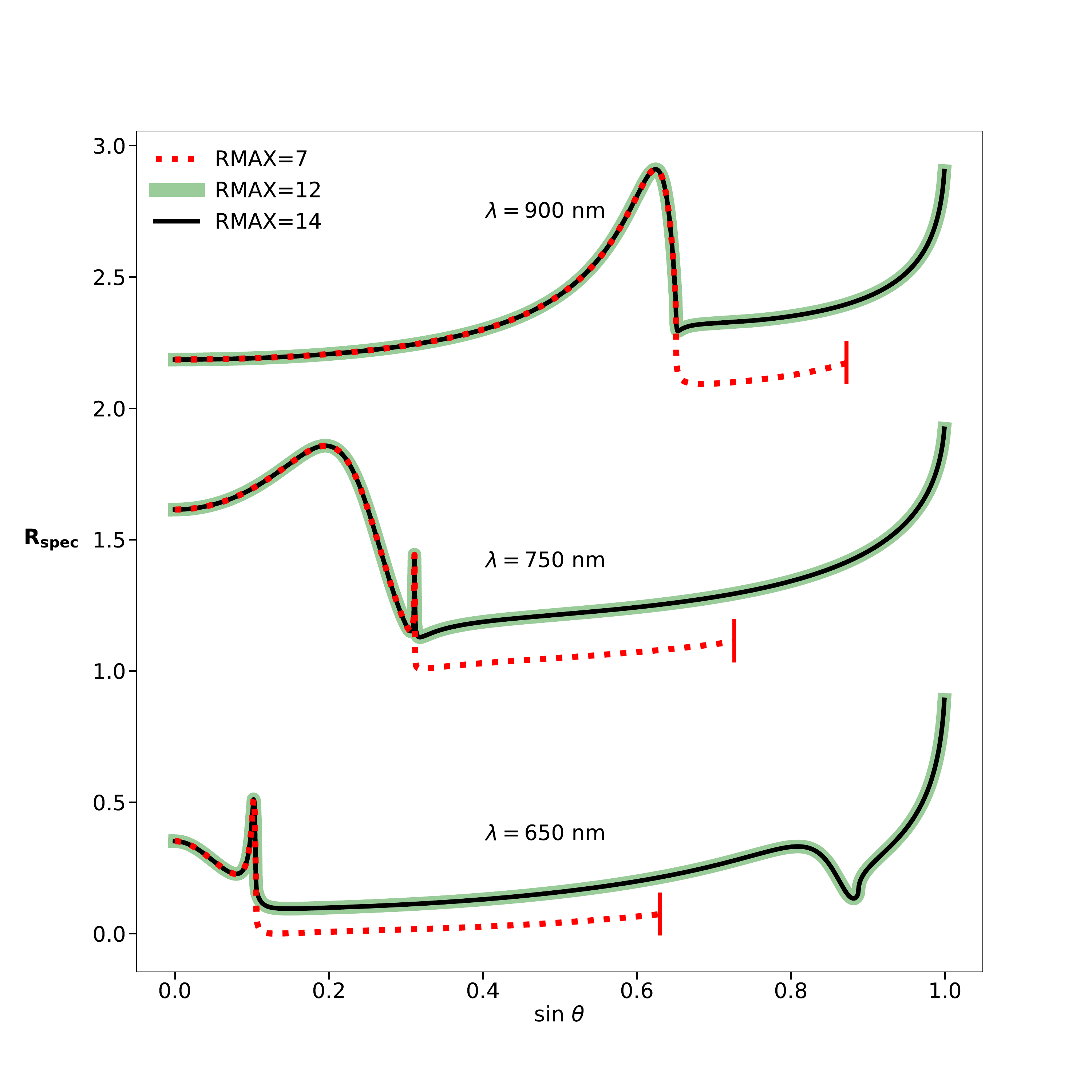}
   \caption{Recalculating specular reflection for TE wave from 2D triangular lattice of Au spheres at different wavelengths of Ref.~\cite{Swiecicki:17}. The spectra of Fig. 4 of Ref.~\cite{Swiecicki:17} were reproduced by selecting LMAX=3. Spectra are shifted by 1 to improve visibility. With increasing incidence angle $\theta$, the in-plane projection ${\bf k}_\parallel$ of the incident wave vector ${\bf k}$ gradually increases. With RMAX as small as seven there is at some point ($\sin\theta=0.63$ at $\lambda=650$ nm, $\sin\theta=0.726$ at $\lambda=750$ nm, and $\sin\theta=0.872$ at $\lambda=900$ nm) not enough of reciprocal vectors to reduce ${\bf k}_\parallel$ into the first Brillouin zone. At this point the plot has been stopped that has been indicated by vertical line. It takes 109 seconds of AMD®Ryzen 5 3500u time to calculate data for the solid line of those plots (1000 data points for each solid line).}
\label{fig:RMAX}
\end{figure}
%%%%%%%%%%%%%%%%

\begin{table}[h!]
 \centering  
 \caption{Varying the cut-off RMAX in Multem 2}
 \vspace*{0.3cm}
 \begin{tabular}{ccc} \toprule
  RMAX  &  IGD(fcc)  &   IGD(sc) \\
  \midrule 
7  & 1 &  5 \\
12 & 7 &  9 \\
14 & 13 &  13 \\
16 &  19 &  21 \\
18 &  19 &  25 \\
19 &  19 &  29 \\
20 &  31 &  37 \\
21 &  31 &  37 \\
22 &  37 &  37 \\
23 &  37 &  45 \\ 
24 &  37 &  45 \\ 
25 &  37 &  45 \\ 
26 &  41 &  57 \\ 
27 &  55 &  61 \\ 
28 &  55 &  61 \\ 
29 &  55 &  69 \\ \bottomrule
\label{tab:RMAX}
\end{tabular}
\end{table}

%%%%%%%%%%%%%%%%%%%%%
\begin{figure}
  \includegraphics[width=\textwidth]{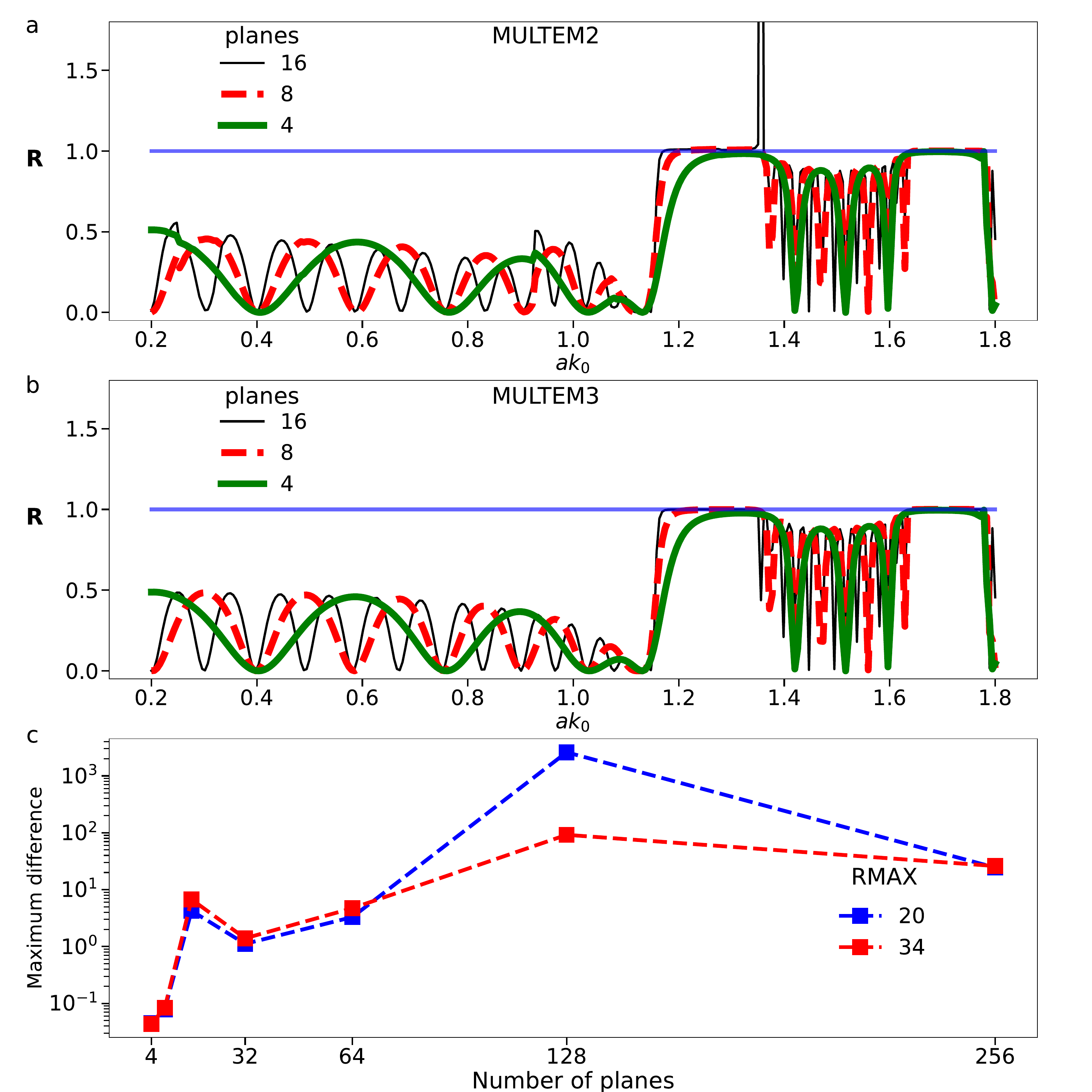}
  \caption{Reflectance spectra of the TE plane wave at the normal incidence to a finite photonic crystal stack of 2D triangular lattices of lossless spheres ($\varepsilon_S = 25$) with radius $0.48a$, where $a$ is the 2D lattice period, stacked along the (111) crystal direction of an fcc lattice. All the spectra were calculated with the cut-off value of LMAX$=7$ and RMAX$=34$. (a) Evolution of the reflectance spectra calculated by Multem 2 (a) and Multem 3 (b) with an increase with the number of planes of spheres parallel to the (111) surface. Whereas the reflectance spectra calculated by Multem 2 occasionally converge to nonphysical results (calculated R exceeds physical bound of $R\le 1$), the converged reflectance spectra calculated by Multem 3 safely obey the physical bound of $R\le 1$ over all range considered and never exceed the horizontal line $R=1$. A noticeable difference is also increasing amplitude of the interference fringes with increasing number of planes in Multem2 just before the first stop gap.
  (c) Maximum absolute difference between Multem 2 and Multem 3 calculations over the frequency range of panels (a) and (b) for different RMAX and the number of planes of spheres stacked along the (111)-direction of an fcc lattice. It takes 5900 seconds of AMD®Ryzen 5 3500u time to calculate data for this figure.}
\label{fig:RMAX2layers}
\end{figure}
%%%%%%%%%%%%%%

RMAX determines via IGD the dimension IGKD (=2*IGD) of scattering matrices QI, QII, QIII, and QIV formed in the routines HOSLAB (for a homogeneous dielectric layer) and PCSLAB (for a plane of spheres). The entries of the scattering matrices QI to QIV are labeled by the respective reciprocal vectors. 
It is worth noting that the transmittance spectra of the incident TE plane wave with $ak_x=0.01$ and $ak_y=0$ in Fig.~\ref{fig:main2} were calculated with RMAX$=16$.
In Fig.~\ref{fig:RMAX}, we recalculated the specular reflection of the incident TE plane wave at different angles of incidence for a 2D triangular lattice of Au spheres 
initially presented in Fig. 4 of Ref.~\cite{Swiecicki:17}
with different cut-off value of RMAX. The spectra of Fig. 4 of Ref.~\cite{Swiecicki:17} were reproduced by selecting LMAX=3.
As the angle of incidence $\theta$ increases, the in-plane projection ${\bf k}_\parallel$ of the incident wave vector ${\bf k}$ in Eq. (\ref{Kkn}) gradually increases. With RMAX as small as seven, there are at some point not enough reciprocal vectors to reduce ${\bf k}_\parallel$ into the first Brillouin zone. At this point the plot was stopped, indicated by the vertical line terminating the dotted red line in Fig.~\ref{fig:RMAX} 
($\sin\theta=0.63$ at $\lambda=650$ nm, $\sin\theta=0.726$ at $\lambda=750$ nm, and $\sin\theta=0.872$ at $\lambda=900$ nm).
Since the convergence parameter RMAX controls the coupling between the layers, it is not surprising that there is no difference between specular reflections with RMAX$=12$ and RMAX$=14$. In fact, all diffraction orders ${\bf k}_n$ (i.e., yielding real $K_{\perp,n}$ in Eq. (\ref{Kkn})) were considered for RMAX$=12$ and RMAX$=14$, respectively, and additional evanescent orders do not play a role in determining the RTA properties of a single plane. 
The purpose of this exercise was to verify that this remains the case numerically and that increasing the size of the scatter matrix does not cause problems.

To demonstrate the effect of evanescent orders on convergence we considered an fcc photonic crystal stack with different number of planes. Each plane consists of lossless spheres ($\varepsilon_S = 25$) arranged into 2D triagonal lattice embedded in a host medium ($\varepsilon_M=1$). Radius of the spheres is $0.48a$, where $a$ is the 2D lattice period. 
The planes were stacked along the (111) crystal direction of an fcc lattice.
We calculated reflectance spectra of the TE plane wave at normal incidence to 
the fcc photonic crystal stack. All the reflectance spectra were calculated with the cut-off value of LMAX$=7$ and RMAX$=32$.
A significant improvement inherent to Multem 3 is evident by comparing Fig.~\ref{fig:RMAX2layers}a and Fig.~\ref{fig:RMAX2layers}b, which demonstrate the reflectance spectra calculated by Multem 2 and Multem 3 with different numbers of planes (4, 8, 16).  
Whereas the reflectance spectra calculated by Multem 2 occasionally converge to nonphysical results (calculated R exceeds physical bound of $R\le 1$), the converged reflectance spectra calculated by Multem 3 safely obey the physical bound of $R\le 1$ over all range considered and never exceed the horizontal line $R=1$.
A noticeable difference is also increasing amplitude of the interference fringes with increasing number of planes in Multem 2 just before the first stop gap appears in Fig.~\ref{fig:RMAX2layers}(a) relative to the same region in Multem 3 in Fig.~\ref{fig:RMAX2layers}(b).

Fig.~\ref{fig:RMAX2layers}c demonstrates the maximum absolute difference between Multem 2 and Multem 3 calculations taken over the  frequency range of the panels (a) and (b) of Fig.~\ref{fig:RMAX2layers} for different RMAX and different number of planes of spheres. Already for four planes, the difference between Multem 2 and Multem 3 can be observed with a naked eye. With increasing the number of planes, the difference between Multem 2 and Multem 3 grows up taking its maximum for 128 planes (Fig.~\ref{fig:RMAX2layers}c).

\subsection{Faddeeva function $w(z)$}
\label{sec:faddeeva}
%%%%%%%%%%%%%%%%%%%%%%%%%%%%%%%%%%%%%%
For lattice sums over 2D lattice in three space dimensions, multems have been implementing powerful, fast, and accurate analytical expressions of Kambe~\cite{Kambe:1967:II,Moroz:2006} in the routine XMAT, originally by Pendry~\cite[Appendix C]{Pendry:LEED:1974}. The first term of the lattice sum, $D_{LM}^{(1)}$, makes use of an incomplete gamma function $\Gamma(a,z)$~\cite{Kambe:1967:II,Moroz:2006}, whose special cases are the error function and the complementary error function~\cite[6.5.17]{AbramowitzStegun},~\cite[7.11.2]{Olver2010}. The required values of the incomplete gamma function $\Gamma(a,z)$ are determined by means of the Faddeeva complex error function~\cite[7.1.3]{AbramowitzStegun},~\cite[7.2.3]{Olver2010},
\begin{equation}
w(z)=e^{-z^2} \left(1+\frac{2i}{\sqrt{\pi}}\int_0^z e^{t^2} \, dt \right) =e^{-z^2} \, \mbox{erfc}(-iz),
\end{equation}
where erfc is the complementary error function~\cite[7.1.3]{AbramowitzStegun}, \cite[7.2.2]{Olver2010}. The values of the Faddeeva function have been supplied by routine CERF, originally by Pendry~\cite[Appendix C, pp. 342-4]{Pendry:LEED:1974}. 
%%%%%%%%%%%%%%
\begin{figure}[h!]
\includegraphics[width=\textwidth]{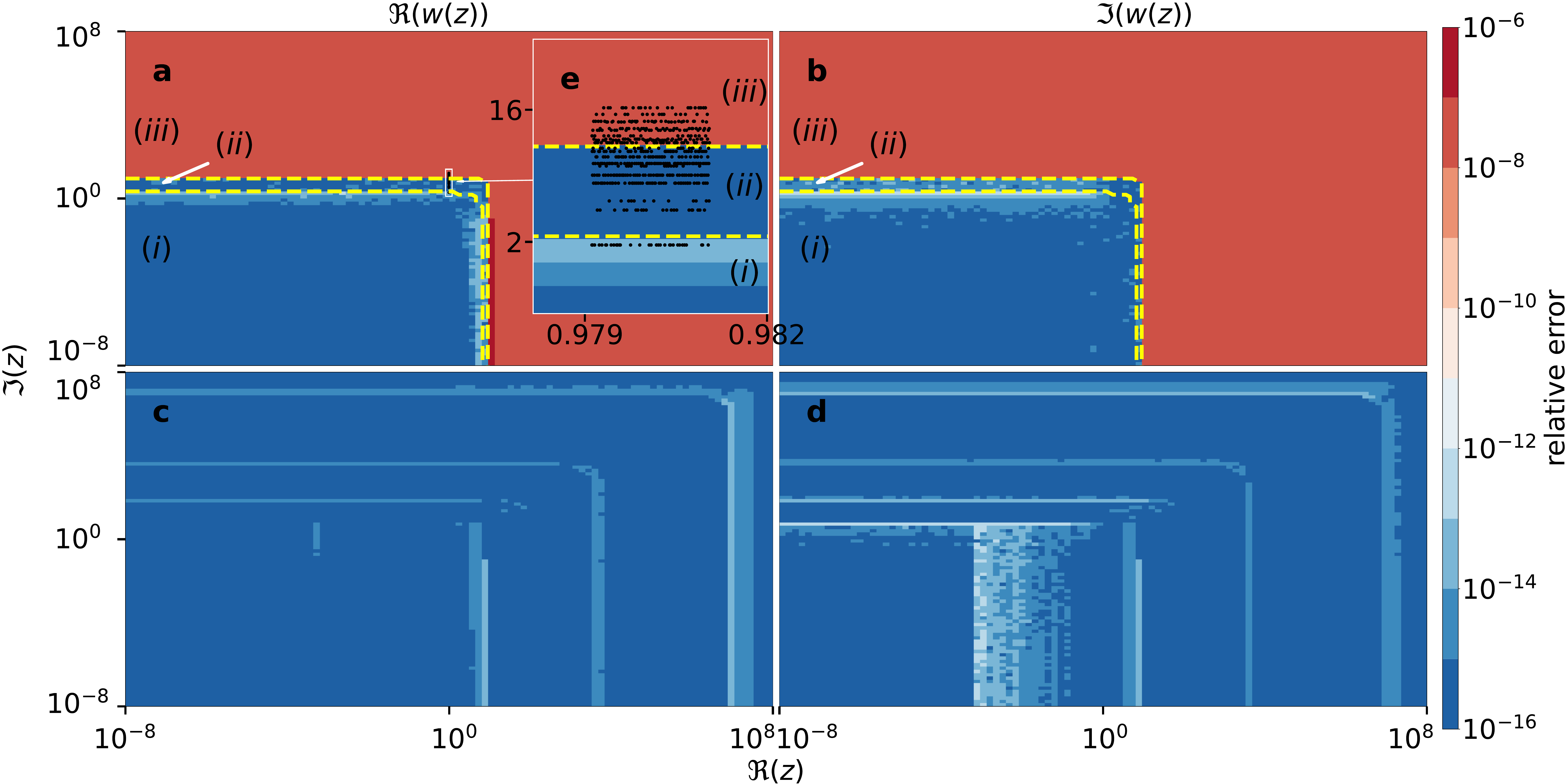}
\caption{ 
%Start with a clause summarizing what is presented in the figure.
(a)-(b) Relative error of calculations of the Faddeeva function, $w(z)$, for complex argument $z=x+iy\, \vert\, x,y \in [10^{-8}, 10^8]$  by the CERF routine of Multem 2. Yellow dashed lines separate  (i) power series, (ii) recurrence based on continued fractions, and (iii) an asymptotic series regions used by CERF to calculate the incomplete gamma function.
 (c)-(d) Relative error of calculations of the Faddeeva function via the state-of-the art Faddeeva function routine of Multem 3.
 The inset (e) shows the complex arguments (indicated by dots) which were needed for the Faddeeva function evaluation in the frequency domain of Fig.~\ref{fig:main2}(a) It takes 17 seconds of AMD®Ryzen 5 3500u time to calculate data for this plot (10000 data points for each map).
 }
\label{fig:faddeeva}
\end{figure}
%%%%%%%%%%%%%%%%%
Depending on the absolute value of its argument $z$, CERF generates the Faddeeva function by one of three methods  (i) a power series ($|z|\le 4$), (ii) an Erd\'elyi recurrence (reproduced in Refs.~\cite[Eqs. (42)-(44)]{Kambe:1967:II} and ~\cite[Eqs. (88)-(89)]{Moroz:2006}) based on continued fractions theory
($4<|z|\le 10$), and (iii) an asymptotic series ($|z|> 10$)~\cite[Sec. 6.5]{AbramowitzStegun}. 
That there may be an issue with the routine has been largely unexpected. CERF is an intrinsic part of great deal of layered multiple-scattering 
~\cite{Pendry:LEED:1974,McL:KKR:1990, McL:KKR1:1990,multem1, multem2,Kafesaki:1999,Kafesaki:2000,Liu:2000} and has been used for decades without any complaints.
%%%%%%%%%%%%%%%
\begin{figure}[htb!]
\includegraphics[width=\textwidth]{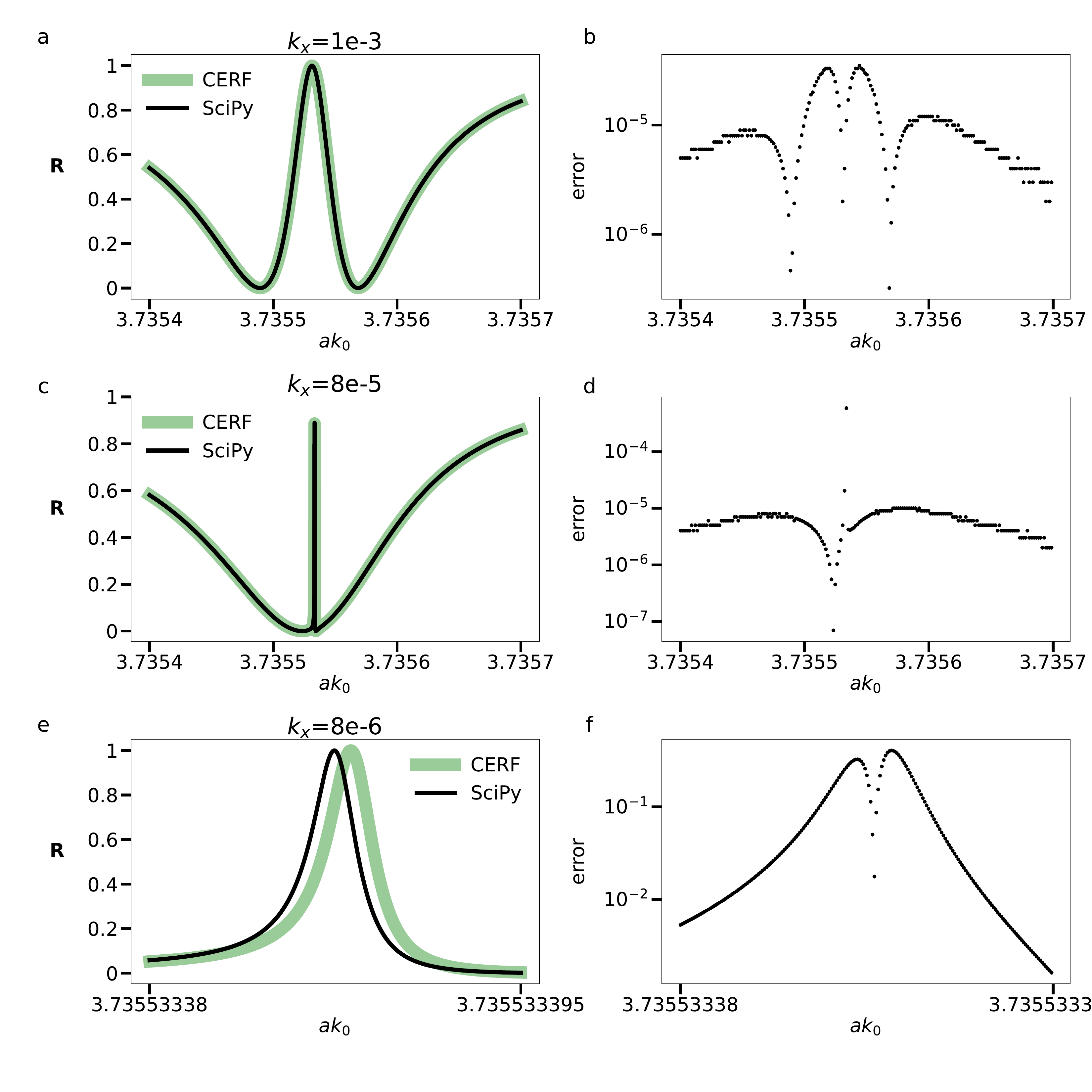}
\caption{(a)-(c)-(e) Reflectance spectra for $k_y=0$ of the incident TE plane wave for a lossless system of Ref.~\cite{Bulgakov:19} 
in the proximity of the BIC's located at the exceptional point at the Brillouin zone point $\Gamma$ (corresponding to $(k_x,k_y)=0$).
The spectra were generated with the cut-off values LMAX=4 and RMAX=16. 
The spectra obtained with the incomplete gamma function $\Gamma(a,z)$ calculated by the original CERF routine are compared to those obtained by SciPy. The narrow peak in panel (c) obtained for R=0.470512 at $ak_0=3.735\ 533\ 387\ 462\ 4623$ by SciPy is characterized by FWHM=$2.46\times 10^{-9}$.
% FWHM=$2.462\ 462\ 234\ 120\  7214\times 10^{-9}$
(b)-(d)-(f) The absolute difference of the calculated spectra. The improvement in the Faddeeva function precision is most pronounced around spectral resonances, which are of the main physical interest. As a rule of thumb, the narrower the resonance, the greater the improvement. It takes 23 seconds of AMD®Ryzen 5 3500u time to calculate data for this plot (1000 data points for each dataset).}
\label{fig:faddeeva_spectra}
\end{figure}

\begin{figure}[htb!]
\includegraphics[width=\textwidth]{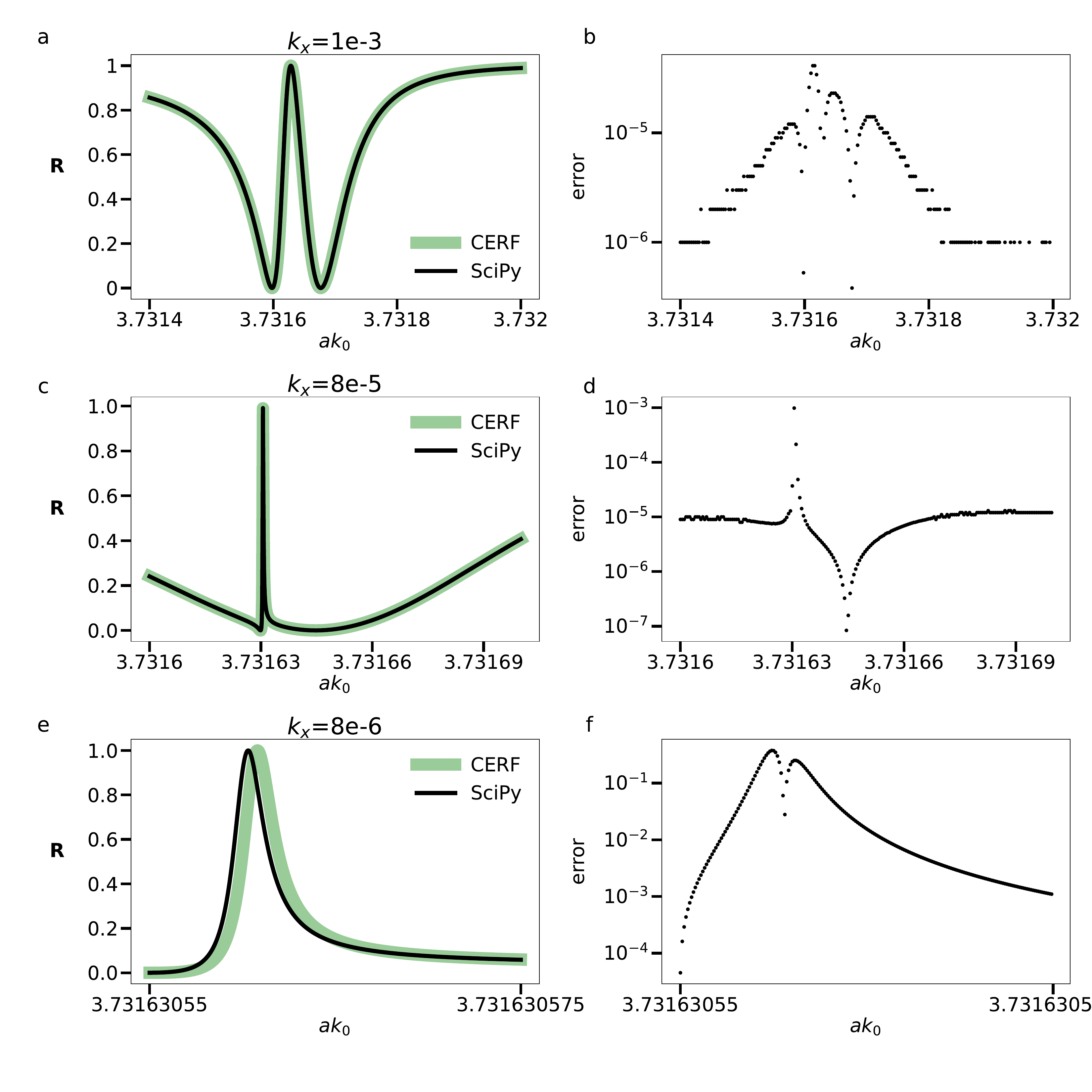}
\caption{Previous Fig. \ref{fig:faddeeva_spectra} recalculated with LMAX=10 and RMAX=16 following LMAX convergence analysis shown in Fig. \ref{fig:main2}(b).
A noticeable difference is that the narrow peak parameters in panel (c) obtained by SciPy are R=0.470665 at $ak_0=3.731\ 630\ 556\ 631\ 632$ is characterized by
FWHM = $3.10\times 10^{-9}$
% FWHM=$3.103\ 102\ 663\ 715\ 1626\times 10^{-9}$ 
and the respective peaks in panel (e) are much closer. This results in different behaviour of error plots in panels b), d), f). It takes 379 seconds of AMD®Ryzen 5 3500u time to calculate data for this plot (1000 data points for each spectrum).}
\label{fig:faddeeva_spectra10}
\end{figure}

There are a number of implementations of Faddeeva function, which were compared in~\cite{FaddeevaImplReview2016}. Multem 3 makes use of C++ code SciPy by Johnson~\cite{ScipyFaddeeva,FaddeevaMIT} to determine the Faddeeva function.
Multem 3 uses Fortran interface~\cite{ArtyomShalevFaddeevaFortran} to Faddeeva implementation mentioned above. The relative error of Faddeeva function calculations was evaluated using mpmath - Python library for real and complex floating-point arithmetic with arbitrary precision~\cite{mpmath}. The values of the Faddeeva function were evaluated with increasing precision until there was no difference at the first 30 significant digits.  
The implementation shows better results with the relative error no more than $1.69\times 10^{-14}$ and $2.67\times 10^{-13}$ for the real and imaginary part of the function values, respectively. Fig.~\ref{fig:faddeeva}(e) shows the complex arguments (indicated by dots) which were needed for the Faddeeva function evaluation in the frequency domain of Fig.~\ref{fig:main2}(a). Thus already for such a basic example one third of the complex arguments of the Faddeeva function were from the asymptotic region with the highest relative error of $4.17 \times 10^{-8}$ (see Fig.~\ref{fig:faddeeva}(a)-(b)). As demonstrated in Figs.~\ref{fig:faddeeva}(c)-(d), Multem 3 features considerably improvement resulting in stable results with relative errors of the Faddeeva function within the range from $1 \times 10^{-16}$ to $7 \times 10^{-16}$, i.e. in the improvement by $8$ orders of magnitude. 
Numerically, the net effect of the above improvement in the accuracy of the Faddeeva function is strongest at spectral resonances, which are of greatest physical interest. As a rule of thumb, the narrower the resonance, the greater the improvement. The latter is one of the reasons why Multem 3 can be safely used to describe BICs.
Fig. \ref{fig:faddeeva_spectra} illustrates this by recalculating the BIC spectra of Ref.~\cite{Bulgakov:19} with LMAX=4, which was the selected cut-off in Ref.~\cite{Bulgakov:19}. Given LMAX convergence analysis shown in Fig. \ref{fig:main2}(b), LMAX=4 used in Ref.~\cite{Bulgakov:19} does not ensure convergence. 
Therefore, Fig. \ref{fig:faddeeva_spectra10} shows the result of recalculating the results presented Fig. \ref{fig:faddeeva_spectra10} with LMAX=10.

\section{Selective multipole expansion}
\label{sec:multipole_expansion}
%%%%%%%%%%%%%%%%%%%%%%%%%%%%%%%%%%%%%
Since the inception of multiple-scattering theory~~\cite{Rayleigh:1892, Darwin:1914:PartI, Darwin:1914:PartII,Pendry:LEED:1974,McL:KKR:1990, McL:KKR1:1990,multem1,multem2}, 
the multipole expansion has been ubiquitous tool for studying the optical properties of resonant scatterers and light-matter interaction~\cite{AlaeeMultipolarDecomposition2019}.
For brevity, only the electric field expansion will be presented in detail below. The magnetic field expansion can be treated analogously. All physical quantities are expressed in conventional SVWF~\cite{Bohren2008} denoted by orbital, $\ell$, and magnetic, $m$, angular numbers, where $\ell\in [1, \infty]$ and $m \in [-\ell, \ell]$. To perform calculations, a limit LMAX for $\ell$ is introduced and convergence for $\ell \in [1, \mbox{LMAX}]$ as a function of LMAX is studied. The incident electric field is expanded as
\begin{equation}
 \label{Einc}
 \mathbf{E}_{i} = \sum_{\ell=1}^{\infty} \sum_{m=-\ell}^{\ell} (a_{\ell m}^{0H}\mathbf{M}_{\ell m}^1 + a_{\ell m}^{0E}\mathbf{N}_{\ell m}^1),
\end{equation}
where $a^{0H(E)}$ are expansion coefficients of the incident field, and $\mathbf{M}^j$ and $\mathbf{N}^j$ are the conventional magnetic and electric SVWF, respectively~\cite{stratton:2007}.
The scattered field is expanded in analogous way,
\begin{equation}
 \label{Escat}
 \mathbf{E}_{s} = \sum_{\ell=1}^{\infty} \sum_{m=-\ell}^{\ell} (a_{\ell m}^{+H}\mathbf{M}_{\ell m}^3 + a_{\ell m}^{+E}\mathbf{N}_{\ell m}^3),
\end{equation}
where the superscript $3$ of the magnetic and electric SVWF's indicates that the spherical Hankel functions~\cite{AbramowitzStegun} are involved.
The T-matrix, with its entries labelled by multi-index $(\ell m)$, connects the expansion coefficients of incident and scattered wave of a single scatterer,
\begin{equation}
 \label{Tmatrix}
 \mathbf{a}^{0P} = \mathbf{T}^P\mathbf{a}^{+P},
\end{equation}
where $\mathbf{a}$ is the vector of coefficients $a_{\ell m}$, $P$ is the multipole type ($E$ for electric and $H$ for magnetic). For spherical scatterers as in multems the T-matrices are necessarily diagonal.
%%%%%%%%%%%%%%%%%%%
\begin{figure}[h!]
\includegraphics[width=\textwidth]{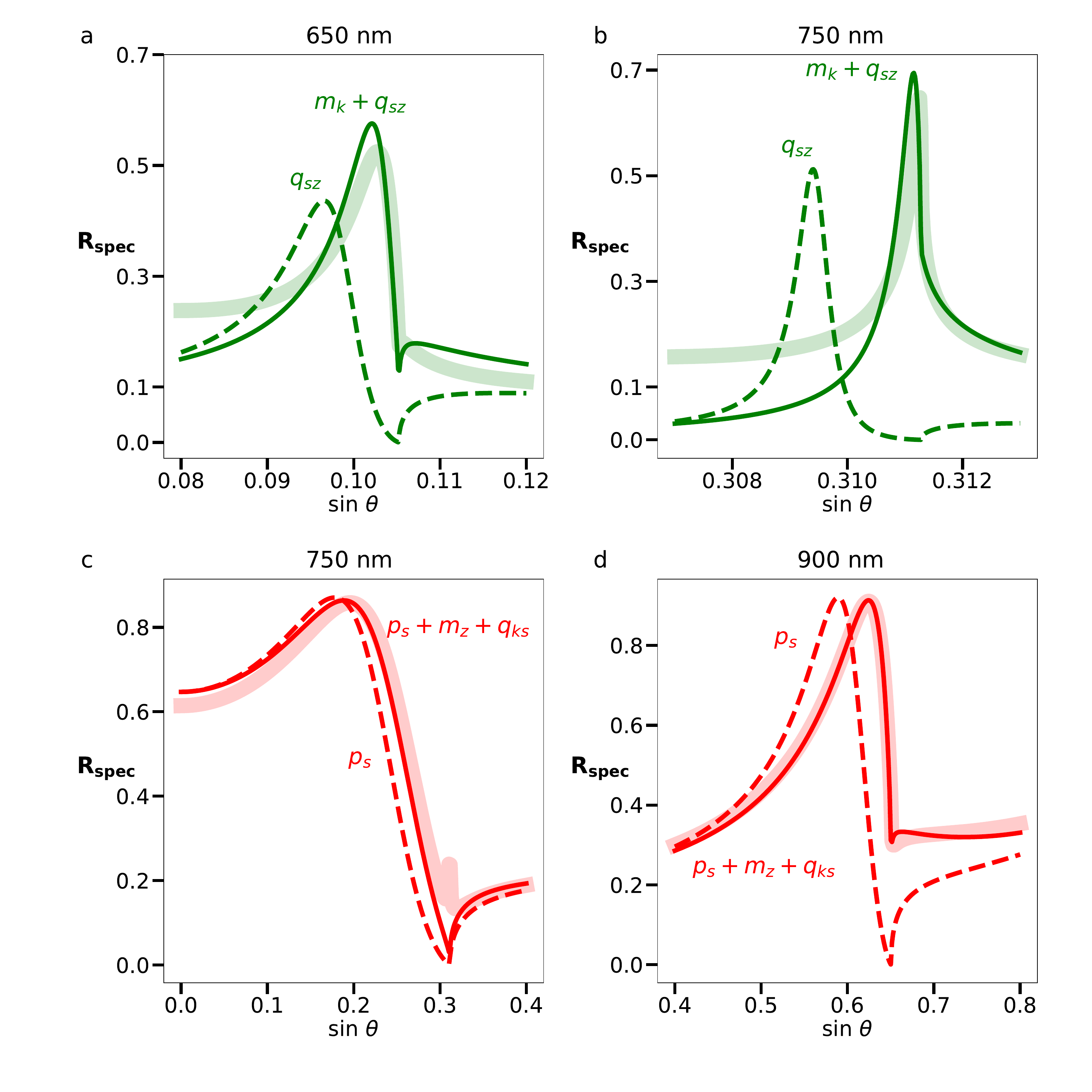}
\caption{Multipole contributions to the specular reflection from triangular lattice of Au-spheres for TE incident wave (recalculated from Ref.~\cite{Swiecicki:17}). 
The spectra of Ref.~\cite{Swiecicki:17} were reproduced by selecting LMAX=2 and RMAX=10. 
Solid semitransparent lines correspond to the total specular reflection. (a)-(b) reproduce the respective left and right panels of Fig. 8 of Ref.~\cite{Swiecicki:17}; (c) reproduces Fig. 5 and the left panel  of Fig. 7 of Ref.~\cite{Swiecicki:17};
(d) reproduces the right panel of Fig. 7 of Ref.~\cite{Swiecicki:17}. The $(p_s,m_z,q_{\kappa s})$ notation of Ref.~\cite{Swiecicki:17} is summarized in the main text. It takes 36 seconds of AMD®Ryzen 5 3500u time to calculate data for this plot (300 data points for each spectrum).}
\label{fig:verification}
\end{figure}
%%%%%%%%%%%%%%%%%%%

In each scattering plane, interaction between spheres arranged on a 2D lattice is taken into account by means of the $\mathbf{\Omega}$ matrices~\cite{multem1}. Multems define coefficients of the scattered field $\mathbf{b}^{+P}$ by solving the following system of linear equations,
\begin{equation}
\label{eq:main}
\begin{pmatrix} \mathbf{I} - \mathbf{T}^E\mathbf{\Omega}^{EE} & \mathbf{T}^E\mathbf{\Omega}^{EH}\\
\mathbf{T}^H\mathbf{\Omega}^{HE} & \mathbf{I} - \mathbf{T}^H\mathbf{\Omega}^{HH} \end{pmatrix}
\begin{pmatrix} \mathbf{b}^{+E}\\ \mathbf{b}^{+H} \end{pmatrix}
 =
 \begin{pmatrix} \mathbf{T}^E\mathbf{a}^{0E}\\ \mathbf{T}^H\mathbf{a}^{0H} \end{pmatrix}\:,
\end{equation}
where $\mathbf{I}$ is the $\mbox{LMAX}\star (\mbox{LMAX} + 2) \times \mbox{LMAX}\star(\mbox{LMAX} + 2)$ unit square matrix.
\begin{comment}
where $\mathbf{I}$ is the $l_{max}(l_{max} + 2) \times l_{max}(l_{max} + 2)$ unit matrix and  $\mathbf{\Omega}^{PP\textquotesingle}$ are the coupling matrices of the scattered field coefficients from a sphere in the 2D lattice $b^{+P}_{lm}$ and the scattered field coefficients from all others spheres in lattice $b^{\textquotesingle P}_{lm}$ (see details in~\cite{multem1} eq.36). \mathbf{!not sure about coupling!}
\end{comment}
In multems the subroutines SETUP and PCSLAB are used to solve the systems of linear equations. The subroutine SETUP constructs the square matrix on the left-hand side of Eq.~(\ref{eq:main}). The subroutine PCSLAB solves Eq.~(\ref{eq:main}) and computes the transmission and reflection matrices for a plane of spheres embedded in a homogeneous host medium.

The principal way to realize multipole expansion in Multem is to process terms where T-matrix is included in both the left- and right-hand sides of Eq.~(\ref{eq:main}). It means that by nullifying terms corresponding to the undesired multipole combination, described by $P$, $l$, and $m$, one can simulate the system with particular multipolar response. 
For verifying multipole expansion regime we used the results of Ref.~\cite{Swiecicki:17}, where a multipolar model of surface-lattice resonances was presented for a triangular lattice of gold spheres. We consider different multipole contributions to reflection coefficient, which depends on the angle of incidence.
Fig.~\ref{fig:verification} shows the exact coincidence with~\cite[Fig. 7 and Fig. 8]{Swiecicki:17}. In the notation of Ref.~\cite{Swiecicki:17}, the parameters $(p_s,m_z,q_{\kappa s},m_{\kappa},q_{sz})$ correspond to the electric dipole moment {\bf p},
magnetic dipole moment {\bf m}, and electric quadrupole moment
{\bf q} in Cartesian basis that are induced on the sphere at a given  in the lattice site by the TE incident plane wave. 
Consequently, the spectra of Ref.~\cite{Swiecicki:17} were reproduced by selecting LMAX=2 and RMAX=10. 
The subscripts $s, \kappa, z$ correspond to the projections along electric field, in-plane wave vector, and normal to lattice respectively. Since Multem's selective multipole expansion works in spherical coordinates, we used spherical-Cartesian relations from~\cite[Appendix C, D and F]{Alaee:2018}.  

Multipole expansion regime allows to carry out simulations in single multipole approximation and has already found its application in simulating BICs in multipolar lattices -- 2D periodic arrays of resonant multipoles~\cite{Gladyshev:2022}.

\section{Limitations and following improvements}
\label{sec:limitations}
%%%%%%%%%%%%%%%%%%%%%%%%%%%%%%%%%%%%%%%%%%%%%%%%%
In our work we have focused on the improvements involving the core part of Multem 2, which is common also to Multem 2 extensions to multiple-scattering of acoustic~\cite{Kafesaki:1999} and elastic~\cite{Kafesaki:2000,Liu:2000} waves, and to the original layer Kohn-Korringa-Rostocker (LKKR) code~\cite{McL:KKR:1990, McL:KKR1:1990}. We did not lift here the well-known limitations of multems that they can work only with the layers of {\em spheres} arranged into {\em simple} Bravais lattice. These two limitations are at present the most significant in separating Multem from simulating arbitrary metamaterial, metasurface, or a photonic crystal designs by commercial software. 
Removing the above limitations is feasible. Non-spherical particles can be treated by replacing TMTRXN routine of multems with the T-matrix code of Mishchenko et al~\cite{MISHCHENKO:2000}. Calculation in this direction with sphere cut by a plane were performed in Sec. 5 of Ref.~\cite{Ederra:2003}.
Complex lattices have been implemented in the electronic version of the original layer Kohn-Korringa-Rostocker (LKKR) code~\cite{ McL:KKR1:1990}
and in the bulk KKR method~\cite{Moroz2002}. It suffices to upgrade the simple Bravais lattice summation routine with that of the LKKR.
Some calculations with Multem 2 upgraded for complex lattice were performed in Refs.~\cite{Moroz2002, Vermolen:2009}.
Our goal is to make the above improvements public in an ensuing upgrade of Multem 3.

\section{Conclusion}
\label{sec:concl}
%%%%%%%%%%%%%%%%%%%%%%%
An updated and revised Multem 3 has been presented that considerably improves its predecessor Multem 2. Multem 3, has been updated to Fortran 2018, with the source code being divided into modules. Multem 3 is equipped with LAPACK, the state-of-the art Faddeeva complex error function routine, and the Bessel function package AMOS. The amendments significantly improve both the speed and precision of Multem 2. Increased stability allows to freely increase the cut-off value LMAX on the number of spherical vector wave functions and the cut-off value RMAX controlling the maximal length of reciprocal vectors taken into consideration. 
An immediate bonus is that Multem 3 can be reliably used to describe bound states in the continuum (BICs). An important message is that much larger values of convergence parameters LMAX and RMAX seem to be required to ensure convergence of the layer coupling scheme than those reported in numerous published work in the past using Multem 2.
We hope that Multem 3 will become a reliable and fast alternative to generic commercial software such as COMSOL Multiphysics, CST Microwave Studio, or Ansys HFSS, and that it will become the code of choice for a large number of research groups for various optimization tasks.

The technical improvements concern the core part of Multem 2, which is common to the extensions of Multem 2 for multiple-scattering of acoustic~\cite{Kafesaki:1999} and elastic~\cite{Kafesaki:2000,Liu:2000} waves, as well as to the original Layer-Kohn-Korringa-Rostocker (LKKR) code~\cite{McL:KKR:1990, McL:KKR1:1990}.
Therefore, the improvements presented here can be readily applied to the above codes as well.

\section{CRediT authorship contribution statement}
%%%%%%%%%%%%%%%%%%%%%%%%%%%%%%%%%%%%%%%%%%%%%%%%%%%%%
\noindent \textbf{Artem Shalev:} Writing - Original Draft, Investigation, Visualization, Software

\noindent\textbf{Konstantin Ladutenko:} Software, Conceptualization, Methodology, Supervision

\noindent\textbf{Igor Lobanov:} Supervision, Writing - Review \& Editing

\noindent\textbf{Vassilios Yannopapas:} Writing - Review \& Editing

\noindent\textbf{Alexander Moroz:} Conceptualization, Writing - Review \& Editing, Validation, Project administration

\section{Conflict of interest}
%%%%%%%%%%%%%%%%%%%%%%%%%%%%%%%
The authors declare no conflicts of interest.

\section{Acknowledgments}
%%%%%%%%%%%%%%%%%%%%%%%%%
The work was supported by the Russian Science Foundation (22-11-00153) (code refactoring including LAPACK, Amos and Faddeeva packages).
A.S. acknowledges Priority 2030 Federal Academic Leadership Program.

\section*{Online supplementary material}
%%%%%%%%%%%%%%%%%%%%%%%%%%%%%%%%%%%%%%%%%%%
Multem 3 simulation setups for an fcc lattice of high refractive index dielectric spheres stacked along the (111) crystal direction and a triangular lattice of Au spheres. Each setup is provided with a Multem 2 compatible input and corresponding output file.

\newpage

\bibliography{references.bib}
\bibliographystyle{unsrt}

%% Authors are advised to submit their bibtex database files. They are
%% requested to list a bibtex style file in the manuscript if they do
%% not want to use elsarticle-num.bst.

%% References without bibTeX database:

% \begin{thebibliography}{00}

%% \bibitem must have the following form:
%%\bibitem{key}...
%%

% \bibitem{}

% \end{thebibliography}

\end{document}

\maketitle

\section{Multem 3 simulation setup for an fcc lattice of high refractive index dielectric spheres stacked along the (111) crystal direction. This setup used to calculate reflectance spectrum for 4 planes has ben used for Fig. 6c of the manuscript}
\label{app:multem3_setup_Fig6}

\subsection{Multem 2 compatible input}
\begin{lstlisting}
           ********************************************
           ********INPUT FILE FOR TRANSMISSION*********
           ********************************************
   KTYPE = 2   KSCAN = 1   KEMB  = 0    LMAX = 7   NCOMP =1   NUNIT = 1
 ALPHA =    1.000000  BETA =    1.000000   FAB =   60.000000  RMAX =  20.000000
  NP =  9  ZINF =  0.200000000000000  ZSUP =  1.800000000000000
  THETA/AK(1) =  0.000000000000000     FI/AK(2) =  0.000000000000000   POLAR =S     FEIN =   0.00

Give information for the "NCOMP" components 

     IT  = 2
     MUMED =   1.00000000   0.00000000     EPSMED=   1.00000000   0.00000000
   NPLAN = 1  NLAYER = 3
       S =   0.48000000     MUSPH =   1.00000000   0.00000000     EPSSPH=    25.000000    0.00000000
xyzDL 0.25 0.14 0.41
xyzDR 0.25 0.14 0.41
       S =   0.48000000     MUSPH =   1.00000000   0.00000000     EPSSPH=    25.000000    0.00000000
xyzDL 0.25 0.14 0.41
xyzDR 0.25 0.14 0.41
       S =   0.48000000     MUSPH =   1.00000000   0.00000000     EPSSPH=    25.000000    0.00000000
xyzDL 0.25 0.14 0.41
xyzDR 0.25 0.14 0.41
\end{lstlisting}

\subsection{multipoles input}
\begin{lstlisting}
[selectors]
is_multipole_type_selected = 0
is_multipole_order_selected = 0
is_m_projection_selected = 0

[regime]
multipole_type = 1
multipole_order = 1
m_projection = 1
\end{lstlisting}

\subsection{Multem 3 output}
\begin{lstlisting}
         ****************************************************
     *** output: transmittance/reflectance/absorbance ***
     ****************************************************
   k_parallel=    0.000000    0.000000     S polarization
   component nr. 1   type:  photonic crystal 
   mu :   1.00000   0.00000 |       1.00000   0.00000
   eps:   1.00000   0.00000 |      25.00000   0.00000
     s:                             0.48000
                                  4 unit layers
   the sample consists of      1 unit slices

             primitive lattice vectors
             ar1 = (      1.0000      0.0000)
             ar2 = (      0.5000      0.8660)
             unit vectors in reciprocal space:
             b1  = (     -0.0000      7.2552)
             b2  = (     -6.2832      3.6276)

   reciprocal vectors     length
  1        0    0       0.000000E+00
  2       -1    0       0.725520E+01
  3       -1    1       0.725520E+01
  4        1    0       0.725520E+01
  5        1   -1       0.725520E+01
  6        0   -1       0.725520E+01
  7        0    1       0.725520E+01
  8        2   -1       0.125664E+02
  9       -2    1       0.125664E+02
 10        1   -2       0.125664E+02
 11       -1    2       0.125664E+02
 12        1    1       0.125664E+02
 13       -1   -1       0.125664E+02
 14        2   -2       0.145104E+02
 15       -2    2       0.145104E+02
 16       -2    0       0.145104E+02
 17        2    0       0.145104E+02
 18        0   -2       0.145104E+02
 19        0    2       0.145104E+02
 20       -3    1       0.191954E+02
 21       -3    2       0.191954E+02
 22       -1    3       0.191954E+02
 23       -2    3       0.191954E+02
 24       -2   -1       0.191954E+02
 25       -1   -2       0.191954E+02
 26        3   -2       0.191954E+02
 27        2   -3       0.191954E+02
 28        1    2       0.191954E+02
 29        1   -3       0.191954E+02
 30        3   -1       0.191954E+02
 31        2    1       0.191954E+02


    frequency   transmittance  reflectance   absorbance
------------------------------------------------------------
  0.200000E+00  0.512189E+00  0.487811E+00  0.272005E-14
  0.400000E+00  0.999445E+00  0.555244E-03 -0.136405E-13
  0.600000E+00  0.544507E+00  0.455493E+00 -0.210942E-14
  0.800000E+00  0.901468E+00  0.985319E-01 -0.148492E-13
  0.100000E+01  0.936466E+00  0.635341E-01 -0.145162E-13
  0.120000E+01  0.205721E+00  0.794279E+00 -0.321965E-14
  0.140000E+01  0.180626E+00  0.819374E+00 -0.190958E-13
  0.160000E+01  0.871686E+00  0.128314E+00 -0.134615E-13
  0.180000E+01  0.955780E+00  0.442205E-01 -0.353328E-13
\end{lstlisting}

\section{Multem 3 simulation setup for triangular lattice of Au spheres. This setup has been used to calculate one point at $\sin\theta=0.08$ for $m_{k}+q_{sz}$ contribution of Fig. 10a of the manuscript}
\label{app:multem3_setup_Fig10}

\subsection{Multem 2 compatible input}
\begin{lstlisting}
                       ********************************************
           ********INPUT FILE FOR TRANSMISSION*********
           ********************************************
   KTYPE = 1   KSCAN = 2   KEMB  = 0    LMAX = 2   NCOMP = 1   NUNIT = 1
 ALPHA =    1.000000  BETA =    1.000000   FAB =   60.000000  RMAX =  10.000000
  NP =   2  ZINF =  1.368421052631579  ZSUP =  1.368422052631579
  THETA/AK(1) =  4.588565735785835     FI/AK(2) =  0.000000000000000   POLAR =S     FEIN =   0.00

Give information for the "NCOMP" components 

     IT  = 2
     MUMED =   1.00000000   0.00000000     EPSMED=      2.102500      0.000000
   NPLAN = 1  NLAYER = 1
       S =   0.21052632     MUSPH =   1.00000000   0.00000000     EPSSPH=   -12.953000      1.120900
xyzDL 0.0  0.0  0.0
xyzDR 0.0  0.0  1.0
     MUEMBL=   1.00000000   0.00000000    EPSEMBL=   1.00000000   0.00000000
     MUEMBR=   1.00000000   0.00000000    EPSEMBR=   1.00000000   0.00000000
\end{lstlisting}

\subsection{multipoles input}
\begin{lstlisting}
[selectors]
is_multipole_type_selected = 1
is_multipole_order_selected = 1
is_m_projection_selected = 1

[regime]
multipole_type = 0 0 1 1
multipole_order = 2 2 1 1
m_projection = -1 1 1 -1
\end{lstlisting}

\subsection{Multem 3 output}
\begin{lstlisting}
        ****************************************************
     *** output: transmittance/reflectance/absorbance ***
     ****************************************************
    angles of incidence (in rad):  theta=   4.59   fi=   0.00     S polarization
   component nr. 1   type:  photonic crystal 
   mu :   1.00000   0.00000 |       1.00000   0.00000
   eps:   2.10250   0.00000 |     -12.95300   1.12090
     s:                             0.21053
                                  1 unit layers
   the sample consists of      1 unit slices
￼
             primitive lattice vectors
             ar1 = (      1.0000      0.0000)
             ar2 = (      0.5000      0.8660)
             unit vectors in reciprocal space:
             b1  = (     -0.0000      7.2552)
             b2  = (     -6.2832      3.6276)
￼
   reciprocal vectors     length
  1        0    0       0.000000E+00
  2       -1    0       0.725520E+01
  3       -1    1       0.725520E+01
  4        1    0       0.725520E+01
  5        1   -1       0.725520E+01
  6        0   -1       0.725520E+01
  7        0    1       0.725520E+01


    wavelength  transmittance  reflectance   absorbance
------------------------------------------------------------
  0.136842E+01  0.718273E+00  0.150384E+00  0.131343E+00
  0.136842E+01  0.718284E+00  0.150378E+00  0.131338E+00
\end{lstlisting}